\documentclass[aps,pre,twocolumn,superscriptaddress,nofootinbib]{revtex4}

\usepackage{amsmath,amssymb}
\usepackage[pdftex]{graphicx}
\usepackage{epstopdf}
\usepackage{hyperref}

\usepackage{color}
\usepackage{newfloat}
\usepackage[]{extraplaceins}
\DeclareFloatingEnvironment[name={SUPP. FIG.}]{supfigure}

\begin{document}

\title{Entangling mobility and interactions in social media}

\author{Przemyslaw A. Grabowicz}\email{pms@ifisc.uib-csic.es}\affiliation{Instituto de F\'{\i}sica Interdisciplinar y Sistemas Complejos IFISC (CSIC-UIB), 07122 Palma de Mallorca, Spain}
\author{Jos\'e J. Ramasco}\affiliation{Instituto de F\'{\i}sica Interdisciplinar y Sistemas Complejos IFISC (CSIC-UIB), 07122 Palma de Mallorca, Spain}
\author{Bruno Gon\c{c}alves}\affiliation{Aix-Marseille Universit\'e, CNRS, CPT, UMR 7332, 13288 Marseille, France}\affiliation{Universit\'e de Toulon, CNRS, CPT, UMR 7332, 83957 La Garde, France}
\author{V\'{\i}ctor M. Egu\'{\i}luz}\affiliation{Instituto de F\'{\i}sica Interdisciplinar y Sistemas Complejos IFISC (CSIC-UIB), 07122 Palma de Mallorca, Spain}

%Max 500 words
\begin{abstract}
Daily interactions naturally define social circles. Individuals tend to be friends with the people they spend time with and they choose to spend time with their friends, inextricably entangling physical location and social relationships. As a result, it is possible to predict not only someone's location from their friends' locations but also friendship from spatial and temporal co-occurrence. While several models have been developed to separately describe mobility and the evolution of social networks, there is a lack of studies coupling social interactions and mobility. In this work, we introduce a model that bridges this gap by explicitly considering the feedback of mobility on the formation of social ties. Data coming from three online social networks (Twitter, Gowalla and Brightkite) is used for validation. Our model reproduces various topological and physical properties of the networks not captured by models uncoupling mobility and social interactions such as:
%i) \color{green} the distance distribution between connected users, ii) the dependence of the reciprocity on the distance, iii) the variation of the social overlap and the clustering with the distance.
i) the total size of the connected components, ii) the distance distribution between connected users, iii) the dependence of the reciprocity on the distance, iv) the variation of the social overlap and the clustering with the distance.
Besides numerical simulations, a mean-field approach is also used to study analytically the main statistical features of the networks generated by a simplified version of our model. The robustness of the results to changes in the model parameters is explored, finding that a balance between friend visits and long-range random connections is essential to reproduce the geographical features of the empirical networks.
\end{abstract}

\maketitle

\section{Introduction}

The advent of the big data revolution has opened the door to the analysis of massive datasets on all aspects of society. New technologies have made possible the access to unprecedented amount of information on human behavior generated unobtrusively whenever people interact with or through modern technologies such as cell phones, online services, mobile applications, etc. This fact is facilitating the pursuit of a computational approach to the study of problems traditionally associated with social sciences~\cite{lazer09-1}. Over the course of the last few years, it has allowed for the development of greater insights, for instance, into human mobility~\cite{Brockmann2006scaling,Gonzalez2008Understanding,Song2010Modelling}, structure of online social networks~\cite{Kwak2010What,Mislove2008Growth}, cognitive limitations~\cite{Miritello2013Time,goncalves11-2}, information diffusion and social contagion~\cite{Bakshy2012role,Ugander2012Structural,leskovec09-1,lehmann2012,grabowicz2012}, the importance of social groups~\cite{grabowicz2012,Grabowicz2013Distinguishing,ferrara12} or even how political movements raise and develop~\cite{borge11,bailon2011,conover13}.

The relation between physical location and social interactions can be also explored with the new available data. In general, people tend to interact and maintain relations with geographically close peers. A tendency that gets reflected in a decay of the social interaction probability with the physical distance. This effect has been observed, for example, in phone call records \cite{Lambiotte2008Geographical,krings09,pith12} and in online friendships~\cite{Liben-Nowell2005Geographic}. Furthermore, it has been shown that online~\cite{Crandall2010Inferring} social links can be inferred from user co-occurrences in space and time and, likewise, that the location of a person can be predicted from the geographic positions of his or her online friends~\cite{Backstrom2010Find}. Some further aspects of the relation between geography and online social contacts have been studied such as the probability that a link at a given distance closes a triangle~\cite{Liben-Nowell2005Geographic,Lambiotte2008Geographical, Scellato2011Socio}, the connections between users in different
countries~\cite{Takhteyev2012Geography}, the social interactions and mobility in emergency situations~\cite{Lu2012Predictability} or the overlap between users' ego networks and how it decays with the distance~\cite{Volkovich2012length}. Multi-parametric inference methods have been applied to empirical data with the aim of predicting link presence and users' locations~\cite{Wang2011Human, Cho2011Friendship, Sadilek2012Finding}. These works show that the accuracy of link prediction is considerably improved by taking into account the geographical information, and that the accuracy of location prediction is enhanced when the online social links are provided.

The availability of geo-localized information has also allowed for a detailed exploration of human mobility~\cite{Brockmann2006scaling,Gonzalez2008Understanding,balcan09,wang09,brockmann10,pith12,simini12}. The length of displacements between locations was found to follow a broad distribution, well fitted by a power-law decaying function~\cite{Brockmann2006scaling,Gonzalez2008Understanding}. The asymmetry of the travels was studied by considering ellipsoidal boundaries to the average individual displacements and analyzing the scaling of the radius of gyration. Memory effects in the individual displacements was also analyzed, finding that individuals' home and workplace have a considerable impact on their mobility patterns \cite{Song2010Modelling}. These results motivated the introduction of several mobility models with the aim of explaining the features observed in the data~\cite{Song2010Modelling,simini12,jia12,szell12,hasan13}. Despite the supporting evidence \cite{pith12}, most of these models lack a connection between mobility and social interactions \cite{Giannotti2012complexity}.

In this work, we lay a bridge between these two worlds by introducing a model coupling social tie formation and spatial mobility. Preceding models considering network structure and geography are uncoupled \cite{Butts2011Geographical, gonzalez06-1}. Our model simulates the movement of individuals and creates links between them when they are physically close mimicking the effect of face-to-face interactions. We study the model both numerically and analytically and confront its results with empirical data obtained from three online social networks. We show that the model generates more realistic networks than uncoupled models.

\section{The Datasets}

We have collected data from online social networks containing both social links and information about the users' physical positions. The first dataset was obtained from Twitter by means of its API~\cite{api}. We identify over $714,000$ single users, who tweeted using a GPS enabled mobile device during the month of August $2011$~\cite{ratkiewicz11-1}. If those users reported various locations in different tweets, the most recent one is taken for the purpose of the study. The other two datasets contain information referring to the users' location check-ins and the social networks of Gowalla and Brightkite~\cite{Cho2011Friendship}. Both were location-based online social networks, in which users can check-in at their current locations and receive information about services in the area as well as about their friends' positions. Gowalla and Brightkite are no longer active but their data is available online~\cite{stanford}. The main statistical features of our three datasets are displayed in Table~\ref{tab:datasets}.

Social interactions across country borders have particular properties and are affected by political, linguistic or cultural factors. We overcome this difficulty by restricting our analysis to the networks within each country. Intra-country mobility and social contacts account for the large majority of a user activity~\cite{Ugander2011Anatomy, State2013Studying}. For simplicity, we focus on the three major countries with more than one thousand users in each of our datasets: the United States (US), the United Kingdom (UK) and Germany (DE). We have analyzed and modeled other countries and found similar results to the ones presented in this manuscript.

\begin{table}[h]
\begin{center}
\begin{tabular}{c|cc|cc|cc|cc}
$\,$ & \multicolumn{2}{c|}{TOTAL($\times 10^3$)} & \multicolumn{2}{c|}{US($\times 10^3$)} & \multicolumn{2}{c|}{ UK($\times 10^3$)} & \multicolumn{2}{c}{DE($\times 10^3$)} \\
$\,$ & N & L & N & L & N & L & N & L \\
\hline
%\hline
Twitter & $714$ & $15000$ & $132$ & $1100$ & $28$ & $117$ & $3.8$ & $8.5$ \\
%\hline
Gowalla & $196$ & $950$ & $46$ & $350$ & $5.2$ & $20$ & $5.2$ & $30$ \\
%\hline
Brightkite & $58$ & $214$ & $27$ & $167$ & $3.1$ & $10$ & $1.3$ & $7.2$ \\
\end{tabular}
\caption{\textbf{Datasets.} Number of users (nodes) $N$ and of links $L$ of the networks obtained from the different geo-localized datasets for the United States (US), the United Kingdom (UK) and Germany (DE).}
\label{tab:datasets}
\end{center}
\end{table}

\begin{figure}
\centering
 \includegraphics[width=8.3cm]{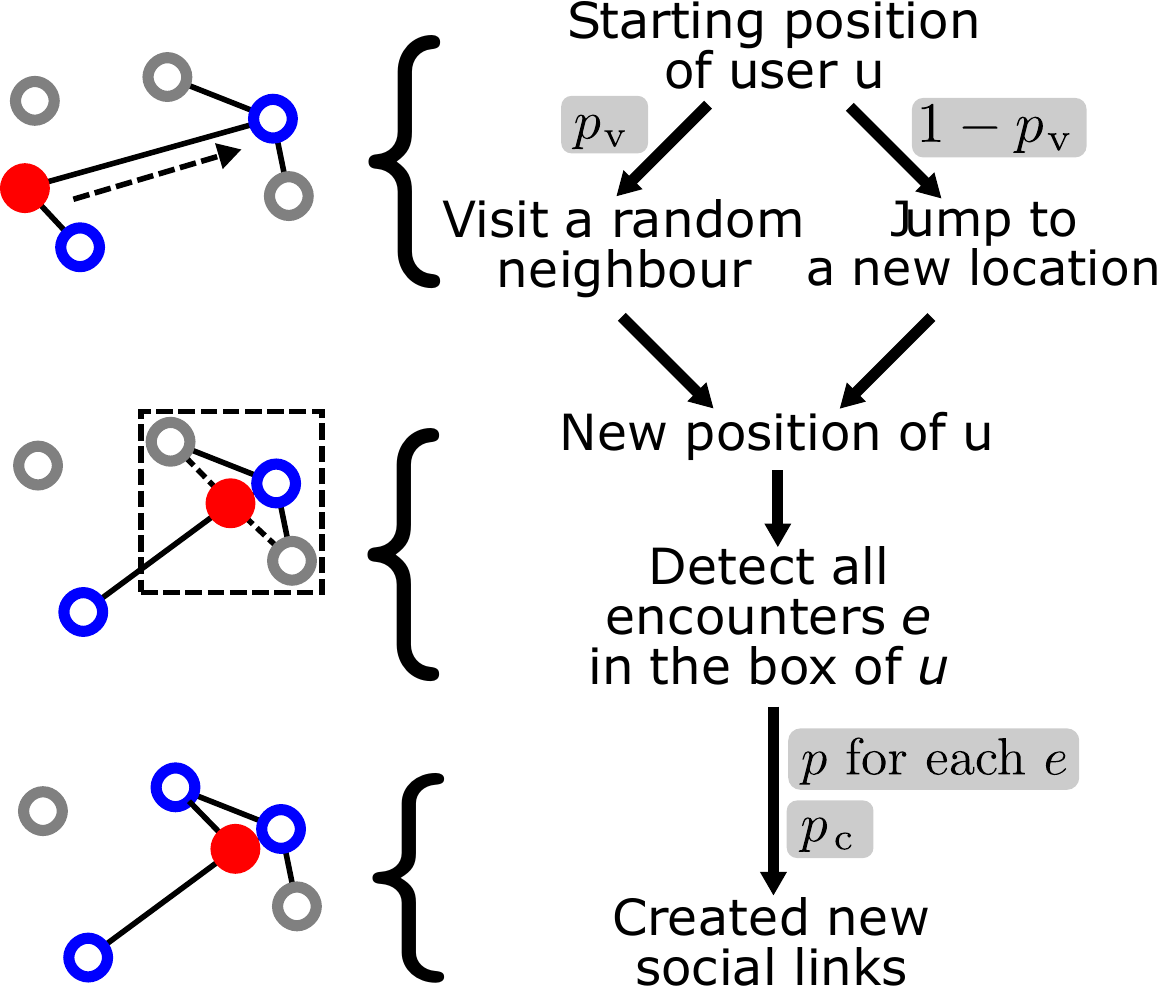}
\caption{
\textbf{Schematic of the TF model.} The central node is the filled red circle and its neighbors are marked in blue. Directionality of links is neglected in this schematic to maintain simplicity.}
\label{fig_schematics}
\end{figure}

\section{The Travel and Friend (TF) model}

The model structure is illustrated in Figure~1. The initial condition is a set of individuals located in the last known positions of the online network users as extracted from the data. At each step of the model, a randomly chosen agent performs actions in two stages:

\begin{enumerate}
\item Travel
 \begin{enumerate}
  \item Visit a randomly selected friend at his current location with probability $p_\text{v}$.
  \item Otherwise, travel to a new location. The distance of travel is obtained from a distribution of jump lengths, while the direction is chosen proportionally to the population density at the target distance.
  \end{enumerate}
\item Friendship
  \begin{enumerate}
  	\item With probability $p$, create directed links to agents within a neighborhood of size $\delta \times \delta$.
  	\item With probability $p_\text{c}$, create a directed connection to a randomly chosen agent anywhere in the system.
  \end{enumerate}
\end{enumerate}

%%%%%%%%%%%%%%%
\begin{figure*}
\centering
\includegraphics[width=17.35cm]{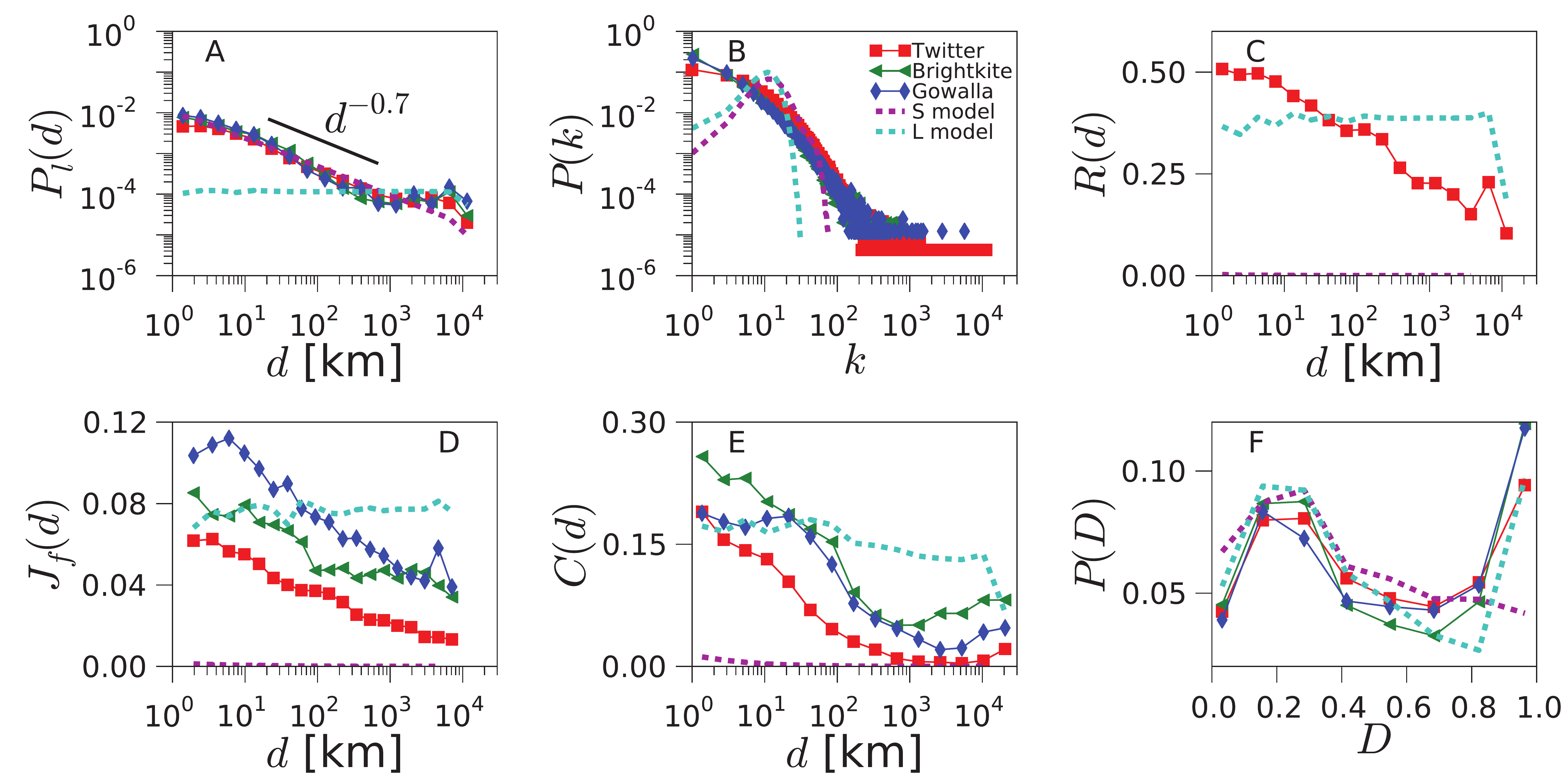}
\caption{\textbf{Network geo-social properties.}
Various statistical network properties are plotted for the data obtained from Twitter (red squares), Gowalla (blue diamonds), Brightkite (green triangles) and the null models (dashed lines), for the US (for the UK and Germany, see Supplementary Figures~\ref{supfig_1} and \ref{supfig_2}). The spatial model (magenta), based on geography, matches well the data in $P_\text{l}(d)$, but yields near-zero values for $R(d)$, $J_\text{f}(d)$ and $C(d)$. The linking model (cyan), based on triadic closure, produces enough clustering, but it does not reproduce the distance dependencies of $P_\text{l}(d)$, $R(d)$, $J_\text{f}(d)$ and $C(d)$.
}
\label{fig_data}
\end{figure*}

The acronym of the TF model comes from the initials of these two stages. The model is iterated until the number of created connections is equal to the number of links measured in the empirical networks. Despite its simplicity, the model incorporates several major features of human behavior. The {\it Travel} stage accounts for both recurring visits to the same location and exploration of new places and the {\it Friendship} component generates both face-to-face contacts and online acquaintances independent of the geography.\footnote{Note that in the Friendship phase both of two possible actions happen concurrently with the respective probabilities.}
The effect of each of the underlying assumptions is systematically explored through analysis of model variants in Appendix B. 

The model has four input parameters: $p_v$, $p$, $p_c$ and $\delta$, besides the distribution of jump lengths. Following the empirical findings of Ref.~\cite{Song2010Modelling}, we take a power-law distribution for the jump lengths with an exponent $-1.55$ for the main simulations shown in this work. Still, other functional shapes for the jump distribution are also discussed in Section~\ref{sec:insights}.
The direction of the jump is chosen proportionally to the population density at the target distance using the gridded population estimates of the world for 2005 with the cell size $2.5'$ \cite{SEDAC2014}.
The values of the probability $p=0.1$ and the box size $\delta=0.001^\circ$ are fixed to match the relation between the probability of friendship and the number of daily spatiotemporal coincidences measured in \cite{Crandall2010Inferring}.\footnote{To this end, we assume that one time step of the model corresponds roughly to one day. Most of our simulations finish in less than a $1,000$ time steps, corresponding to a few years, which is of the order of magnitude of users' lifetime, given that Twitter was founded in 2005 and our dataset is from 2011.} Furthermore, we tested different values of $\delta$ and $p$ and did not observe strong deviations in the model results. This leaves us only with $p_v$ and $p_c$ as free model parameters, we will systematically explore in the coming sections the impact of these parameters on the model results, since, as it will be shown, they are essential for generating network comparable with the empirical ones.

\section{Geo-social properties of the networks}

We start by establishing a set of metrics in order to characterize networks structure and its relation to geography. First, we measure the probability of two users to have a link at a certain distance $P_\text{l}(d)$. It is defined as the ratio between the number of existing links at distance $d$ and the total number of users pairs separated by $d$, and thus it is constrained to the interval $\left[0,1\right]$. $P_\text{l}(d)$ decays slowly with the distance for empirical networks, essentially as a power-law with exponent $-0.7$, which is followed by a plateau for very large distances (see Figure~2A). This functional shape remains identical for all the countries and all the datasets considered. It matches, besides, the behavior reported in the literature for online social systems~\cite{Liben-Nowell2005Geographic,Scellato2011Socio}.

A second metric that we consider is the degree distribution of the social networks (see Figure~2B for the empirical networks). For Twitter, which has a directed social network, we consider the degrees of its symmetrized version. The distribution $P\left(k\right)$ displays heavy tail in all the datasets, even though there are slight differences between them.

Connections in Twitter are directed: one user follows the messages emitted by another. Reciprocated connections indicate mutual interest between the two users and a closer type of social relation~\cite{goncalves11-2,Grabowicz2013Distinguishing}. To assess how geography and reciprocity correlate, we measure the probability $R\left(d\right)$ of reciprocation conditional on a link at a distance $d$ (Figure~2C). We find that the reciprocity decreases with the distance in all the countries analyzed. This trend is consistent with the idea that stronger relations occur close to where users spend most of their time, with some longer connections composed of friends who moved, former residences, online acquaintances, etc. Furthermore, long not-reciprocated connections may include users following public figures or celebrities.

With the aim of quantifying social closeness between users, we define the social overlap $J_\text{f}$ of two connected users $i$ and $j$ as
\begin{equation}
J_\text{f}=\frac
{ \left|\mathcal{K}_{i} \cap \mathcal{K}_{j}\right| }
{ \left|\mathcal{K}_{i} \cup \mathcal{K}_{j}\right|-2 }
\end{equation}
where $\mathcal{K}_{i}$ represents the set of friends of user $i$. $J_\text{f}$ is inspired by the Jaccard index but is modified to ensure that it takes a value of $1$ if $i$ and $j$ share all their friends, and $0$ if they have no common friends. In Figure~2D, the average of the social overlap $J_\text{f}\left(d\right)$ over all pairs of connected users is plotted as a function of the distance between them. The social overlap decreases with the distance. The functional shape of the curves is similar for all the datasets, even though the overlap level is different for each of them. For Twitter, we use the symmetrized version of the network to study social overlap and clustering.

Another well known phenomenon in social networks is triadic closure. As one individual has a close relation with other two persons, there are high chances that these two individuals end up creating a social relation between themselves. In network analysis, a magnitude that quantifies this effect is the average clustering coefficient $C$. It is defined as the ratio between the number of closed triads and the total number of triads in the network. 
%A triad consists of three connected nodes: if they are fully connected, forming a triangle, the triad is closed, otherwise it is open. 
A triad is a sequence of $3$ nodes $i,j,k$ such that the central node $j$ is connected to both extreme nodes $i$ and $k$. A closed triad is a triad that has also an edge between $i$ and $k$, forming a triangle. Note that a triangle consists of $3$ triads centered on different nodes.
The effect of the distance on the clustering coefficient can be incorporated by measuring the distances from each central node $j$ to two neighbors $i$ and $k$ forming a triad, $d = d_{ij}+d_{jk}$, and calculating the network clustering restricted to triads with distance $d$. This new function $C(d)$ is the probability of closing a triangle given the distance $d$ in a triad
\begin{equation}
C(d)=\frac{\Delta(d)}{\Lambda(d)},
\label{eq:clustering_distance}
\end{equation}
where $\Lambda(d)$ and $\Delta(d)$ are the numbers of triads and closed triads for the distance $d$, respectively. The value of the global clustering coefficient $C$ can be recovered by averaging $C(d)$ over $d$. In the datasets, we observe a drop in $C(d)$ followed by a plateau, which is best visible for the US networks (Figure ~2E).

%%%%%%%%%%%%%%%%%%%%%%%%%%%%%%%%%%%%%%%%%
\begin{figure*}
\centering
 \includegraphics[width=17.35cm]{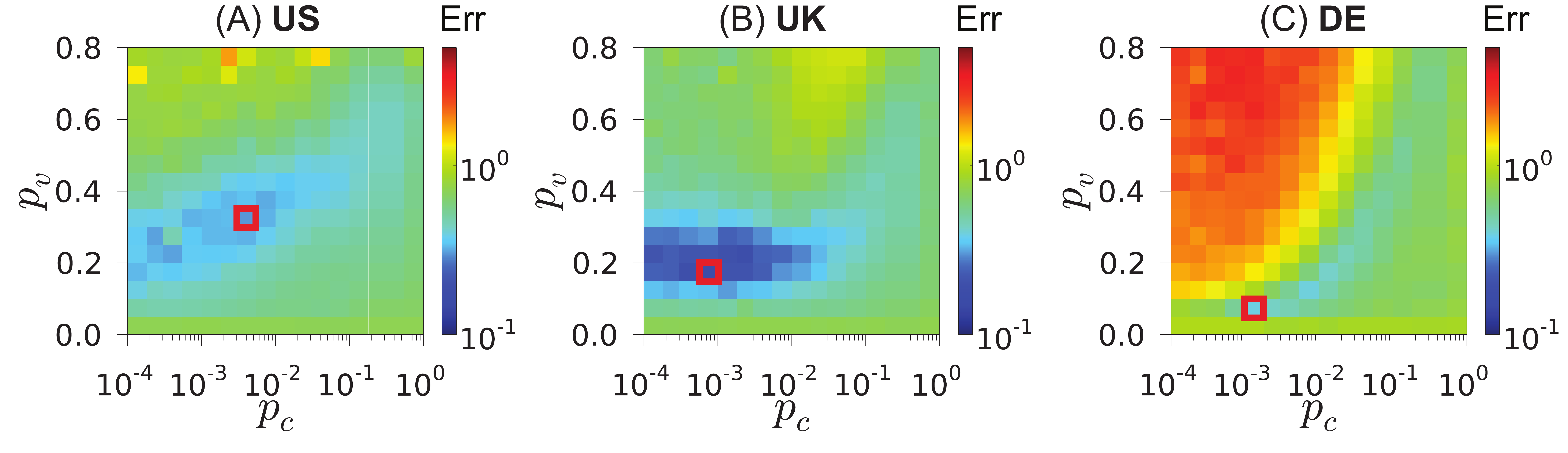}
\caption{
\textbf{Fitting the TF model.} Values of the error $\text{Err}$ when $p_\text{v}$ and $p_\text{c}$ are changed. The minimum error for each of the plots is marked with a red rectangle.}
\label{fig_colormap}
\end{figure*}

Given a triangle, several configurations are possible if there is diversity in the edge lengths. The triangle can be equilateral if all the edges have the same length, isosceles if two have the same length and the other is smaller, etc. We estimate the dominant shapes of the triangles in the network by measuring the disparity $D$ defined as:
\begin{equation}
D= 6\left[\frac{ d_1^2+d_2^2+d_3^2 }{ (d_1+d_2+d_3)^2 }-\frac{1}{3}\right],
\label{eq_disparity}
\end{equation}
where $d_1$, $d_2$ and $d_3$ are the geographical distances between the locations of the users forming the triangle. The disparity takes values between $0$ and $1$ as the shape of the triangle passes from equilateral to isosceles, where one edge is much smaller than the other two. $D$ shows a distribution with two maxima in the online social networks (Figure~2F), for low and high values. The two most common geometries of the triangles are: i) all $3$ users are at a similar distance, ii) $2$ users are close to each other, while the third one is distant. Since most edges correspond to small distances, this means that most triangles are constituted by three users that are all close to each other geographically. However, the stretched isosceles configuration is also relatively common.

%%%%%%%%%%%%%%%%%%%%%%%%%%%%%%%%%%%%%%%%%%%%%%%%%%%%%%%%
\begin{figure*}
\centering
\includegraphics[width=17.35cm]{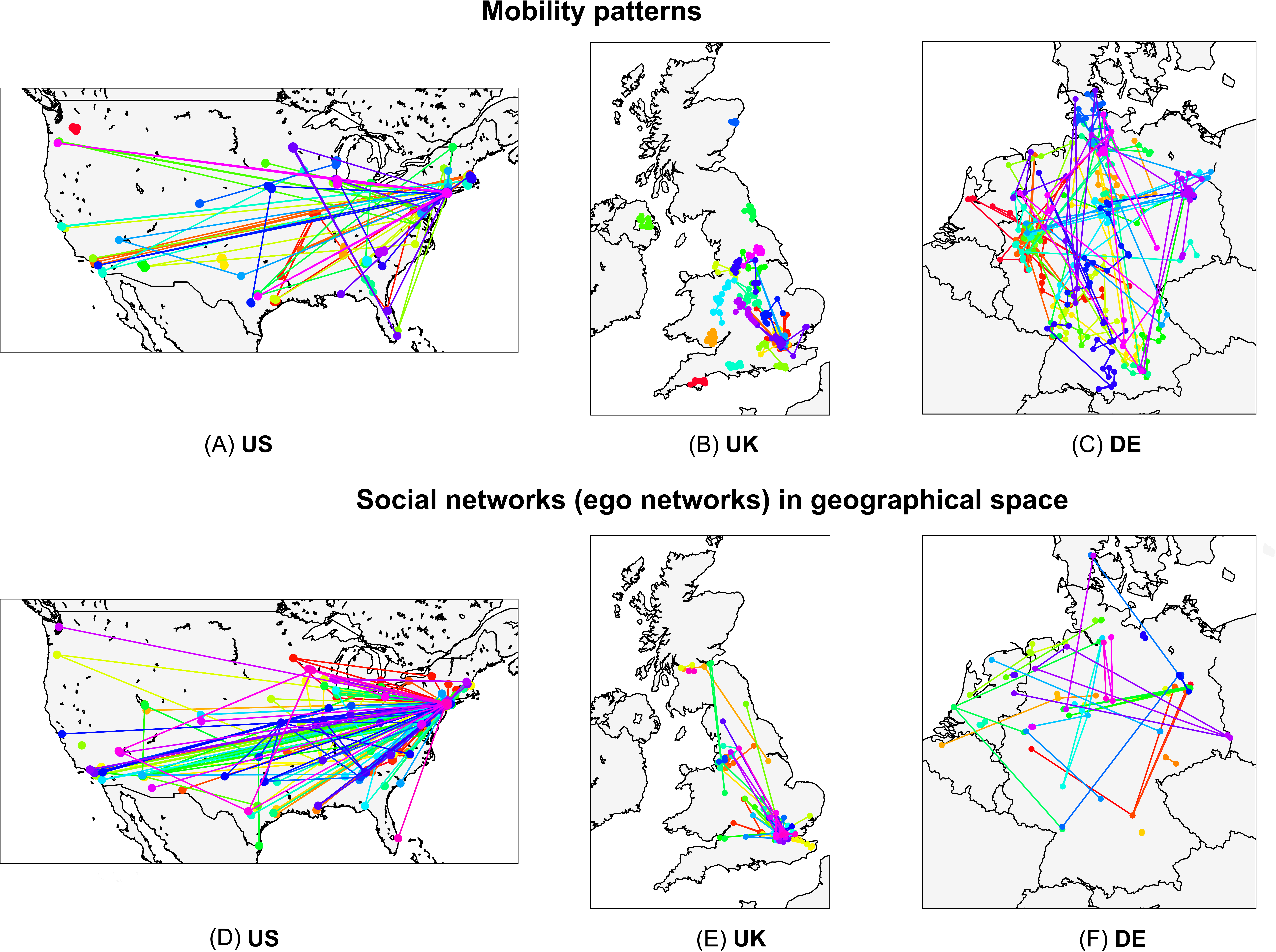}
\caption{\textbf{Simulation results: mobility and social networks.} Mobility (upper row) and ego networks (lower row) of $20$ random users (different colors) for the instances of the TF model yielding the lowest error $\text{Err}$ (see Figure~3). Mobility network shows mobility patterns of individual users throughout entire simulation. Ego network shows the social connections at the end of the simulation.}
\label{fig_realmap}
\end{figure*}

Summarizing, we have defined the following metrics in order to characterize the networks structure and its relation to geographical distance:
\begin{itemize}
\item {\bf $P_\text{l}(d)$}: Probability of linking at a distance $d$ (Figure~2A).
\item {\bf $P(k)$}: Degree distribution (Figure~2B).
\item {\bf $R(d)$}: The probability of reciprocation conditional on a link at a distance (Figure~2C).
\item {\bf $J_\text{f}(d)$}: Average overlap as a function of the distance (Figure~2D).
\item {\bf $C(d)$}: Clustering coefficient as a function of the triad distance (Figure~2E).
\item {\bf $P(D)$}: Distribution of distance disparity for the triangles' edges (Figure~2F).
\end{itemize}
We will use these metrics in the coming sections to estimate the ability of model to produce social networks comparable with those obtained from the empirical datasets.

%%%%%%%%%%%%%%%%%%%%%%%%%%%%%%%%%%%
%\section{Model fitting}
\section{Results}

\subsection{Model calibration}

%%%%%%%%%%%%%%%%%%%%%%%%%%%%%%%%%%%%%%%%%%%%%%%%%%%%%%
\begin{figure*}
\centering
\includegraphics[width=17.35cm]{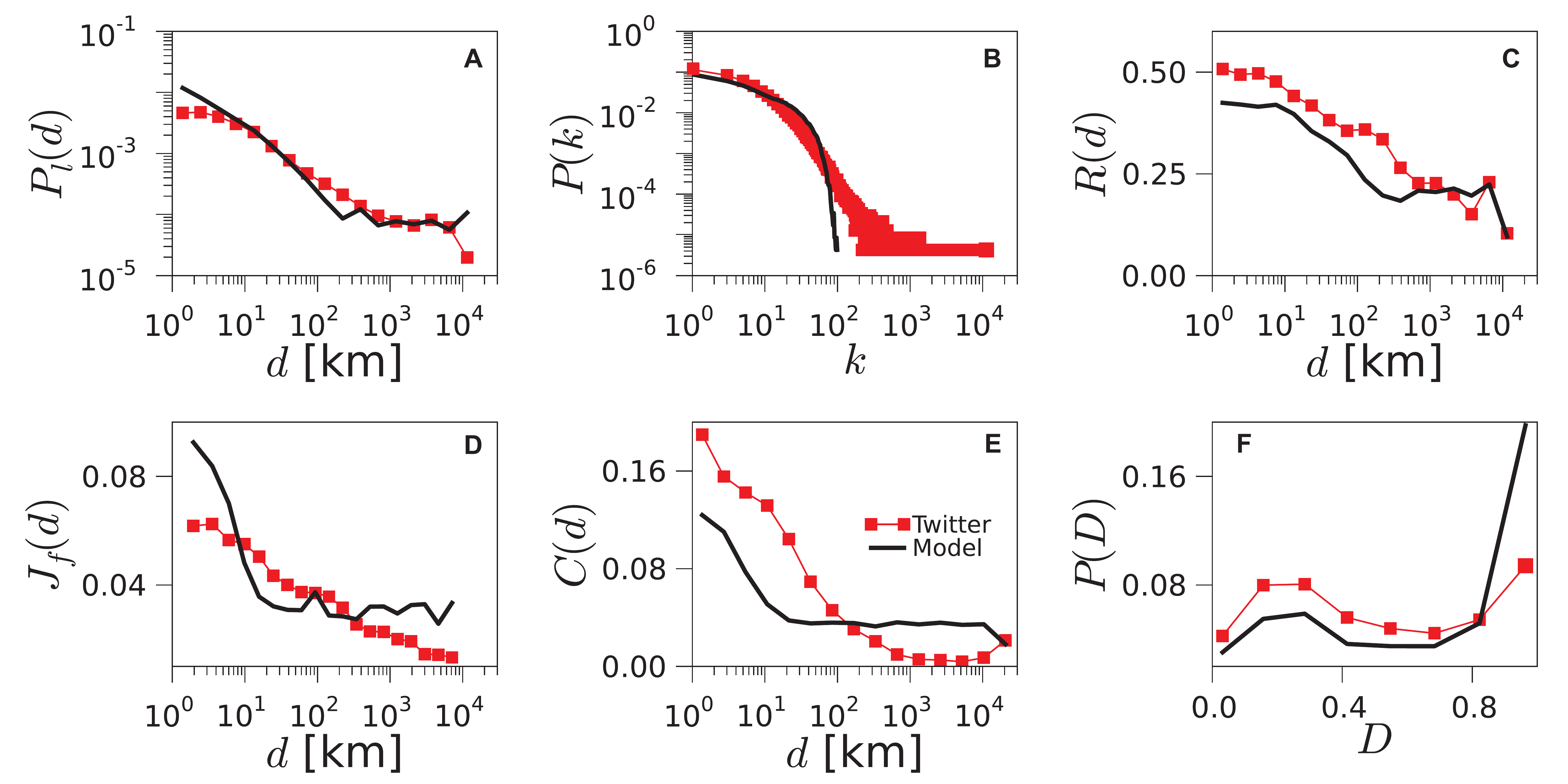}
\caption{\textbf{Geo-social properties of the model networks.} Various statistical properties are plotted for the networks obtained from Twitter data (red squares) and from simulation of the TF model (black line) for the US. Corresponding results for the UK and Germany can be found in Supplementary Figures~\ref{supfig_3} and \ref{supfig_4}.}
\label{fig_main}
\end{figure*}

Next, we will find a compromise between the different metrics and search for the parameter values for which a given model best fits simultaneously the various statistical properties. To do so, we define an overall error $\text{Err}$ to quantify the difference between the networks generated with the model and the empirical ones. The parameters of the model are then explored to find the values that minimize $\text{Err}$. We measure the error $\text{Err}\left[X\right]$ for each property $X$ and take the average over all the properties
\begin{align}
\text{Err}= & \frac{1}{8} \, \Big\{ \text{Err}\left[{P_\text{l}\left(d\right)}\right] + \text{Err}\left[{P\left(k\right)}\right] + \text{Err}\left[{R\left(d\right)}\right] + \text{Err}\left[{J_\text{f}\left(d\right)}\right] \nonumber\\ &
+ \text{Err}\left[{C\left(d\right)}\right] + \text{Err}\left[{P\left(D\right)}\right] + \text{Err}\left[{N_\text{c}}\right] + \text{Err}\left[{C_\text{avg}}\right] \Big\},
\label{eq_toterror}
\end{align}
where $N_\text{c}$ is the total number of nodes in connected components of the network and $C_\text{avg}$ is the undirected local clustering coefficient averaged over the $N_\text{c}$ connected nodes. The local clustering coefficient of a node $i$ is defined as the ratio between number of closed triads centered on node $i$ and the total number of triads centered on that node.

The properties $X$ integrating $\text{Err}$ can be scalars, functions or distributions and encompass different orders of magnitude. We define the error of a property $X$ as
\begin{equation}
\text{Err}\left[X\right]= \frac{ \sum\nolimits_{i=1}^n\left|y_i^X-f_i^X\right| }{\sum\nolimits_{i=1}^n \left|y_i^X\right| },
\label{eq_error}
\end{equation}
where $y_i^X$ is the $i$-th observed value of the property $X$, $f_i^X$ is the corresponding $i$-th value of the property obtained by the model. In the case of a distribution, $i$ runs over the $n$ measured bins, while for a scalar (such as the number of nodes or the clustering coefficient) the sum has only one term.

We perform a Latin square sampling of the parameter space of $p_\text{v}$ and $p_\text{c}$ as shown in Figure~3 in order to find the minimum value of $\text{Err}$. The parameter space is covered uniformly in a linear scale for $p_\text{v}$ and in a logarithmic one for $p_\text{c}$. For all the countries, the minimum value of the error is obtained for $p_\text{v}$ in the interval $\left(0.05,0.3\right)$ and $p_\text{c}$ in the range $(5\cdot 10^{-3},5\cdot 10^{-2})$. The values of $\text{Err}$ found at the minimum are $0.30$ for the US, $0.18$ for the UK and $0.39$ for Germany. For simplicity, we focus on the Twitter networks only, although similar results are obtained for the other datasets.

%%%
\subsection{Simulations for the optimal parameters}

An example with the displacements between the consecutive locations and the ego networks for a sample of individuals, as generated by the TF model, are displayed in Figure~4. The parameters of the model are set to the ones that correspond to the minimum of the error $\text{Err}$. As shown, the agents tend to stay close to their original positions. Occasional long jumps occur due to friend visits that live far apart. In this range of parameters and simulation times, the main mechanism for generating long distance connections is random linking (controlled by $p_\text{c}$). Agents typically return back to their original positions because this is where most of their contacts live. The frequency of the long distance jumps and connections varies for the three countries due to the different spatial distribution of the user populations. In the ego networks, the presence of multiple triangles with long distance edges can be observed.

\begin{table}[tbh!]
\begin{center}
\begin{tabular}{c||c|c|c|c|c|c|c|c||c}
$\,$ & $P_\text{l}$  & $P(k)$  & $R(d)$  & $J_\text{f}(d)$ & $C(d)$ & $P(D)$ & $N_
\text{c}$ & $C_\text{avg}$ & Err \\ 
\hline 
US & $0.60$ & $0.35$ & $0.17$ & $0.33$ & $0.56$ & $0.41$ & $0.01$ & $0.05$ & $0.30$ \\
UK & $0.20$ & $0.43$ & $0.15$ & $0.25$ & $0.25$ & $0.06$ & $0.02$ & $0.34$ & $0.18$ \\
DE & $0.56$ & $0.60$ & $0.53$ & $0.36$ & $0.21$ & $0.36$ & $0.56$ & $0.17$ & $0.39$ \\
\end{tabular} 
\caption{\textbf{The contribution of each of the properties to the total error of the TF model.} Value of the error $\text{Err}\left[X\right]$ per property $X$ at the minimum of the total error $\text{Err}$ for Twitter for the three considered countries.}
\label{tab:partialerr}
\end{center}
\end{table}

The geo-social properties of the networks generated by the TF model are shown in Figure~5 for the US and in Supplementary Figures~\ref{supfig_3} and \ref{supfig_4} for the UK and Germany, respectively. Additionally, we show how each of the introduced properties contributes to the total error of the model in Table~\ref{tab:partialerr}.
The model is able to reproduce the trends in the probability $P_\text{l}\left(d\right)$, the reciprocity $R\left(d\right)$, the social overlap $J_\text{f}\left(d\right)$ and the disparity distribution $P\left(D\right)$ with good accuracy. The difficulties encountered with the degree distribution $P\left(k\right)$ and the clustering as a function of the distance $C\left(d\right)$ are not unexpected since the model does not incorporate mechanisms to explicitly enhance the heterogeneity in the agents' contacts nor favor any specific dependence of the clustering on the distance.\footnote{We have tested variants of the TF model in which connections are created using the preferential attachment rule. The overall fitting error for these variants of the model is not lower than for the basic TF model, as we show in Appendix~I.}

%%%
\subsection{Insights of the TF model}
\label{sec:insights}

In this section we explore two null models uncoupling mobility and social interactions to help us interpret the mechanisms acting in the TF model. The first null model, the spatial model (S model), is based solely on the geography and consists of randomly connecting pair of users with a probability depending on the distance, but does not take network structure into account. The second null model, the linking model (L model), in contrast, is based only on random linking and triadic closure, and it is equivalent to the TF model without the mobility. We consider the two uncoupled null models and compare their results with those of the TF model. In this way, we demonstrate the importance of the coupling through a realistic mobility mechanism to reproduce the empirical networks.

The spatial model (S model) consists of randomly connecting pair of users with a probability that decays as power-law of the distance between them (suggested in~\cite{Butts2011Geographical}). The exponent of the power-law is fixed at $-0.7$ following Figure~2A. The results of the S model are shown in the panels of Figure~2. While it is set to match $P_\text{l}\left(d\right)$, other properties such as $P(k)$, $R\left(d\right)$, $J_\text{f}\left(d\right)$, $C\left(d\right)$ or $P\left(D\right)$ are not well reproduced. The S model fails to account for the high level of clustering and reciprocity in the empirical networks and for their dependence on the distance. The error $\text{Err}$ of this null model is between $0.66$-$0.76$ for the three countries, around twice the error of the TF model (see Figure~6).

The linking model (L model) is a simplified version of the TF model, without random mobility and the box size $\delta \rightarrow 0$. Agents move to visit their contacts with probability $p_\text{v}$, whereas with probability $1-p_\text{v}$ they do not perform any action. In this version of the model, users can connect only by random connections or when two of them coincide, visiting a common friend, which leads to triadic closure. These two processes do not depend on the distances between the users. A thorough description can be obtained with a mean-field approach (see Section~\ref{sec:mf}).
The results of the L model are shown in Figure~2. Due to the triangle closing mechanism, this null model creates networks with a considerable level of clustering. However, it does not reproduce the distance dependencies of $P_\text{l}(d)$, $R(d)$, $J_\text{f}(d)$ and $C(d)$. The error $\text{Err}$ of the L model is also around twice higher than the error of the TF model (see Figure~6).

The geography and the structure are coupled in the TF model through the random mobility. Changes in the underlying mobility mechanism affect the quality of the results. The lowest $\text{Err}$ values are obtained with the power-law distribution in the jump lengths, while normal or uniformly distributed jumps yield worse results (e.g., for the US the TF model has $\text{Err}$ lower by $0.5$ and $1.5$ than the TF-normal and the TF-uniform models, respectively, as shown in Figure~6). 

Simplified models that neglect either geography or network structure perform considerably worse than the TF model in reproducing the properties of real networks. Likewise, non-realistic assumptions on human mobility mechanism yield worse results than the default TF model. To conclude, the coupling of geography and structure through a realistic mobility mechanism produces networks with significantly more realistic geographic and structural properties.

%%%%%%%%%%%%%%%%%%%%%%%%%%%%%%%%%%%%%%%%%%%%%%%%%%%%%%%
\begin{figure}
\centering
\includegraphics[width=8.3cm]{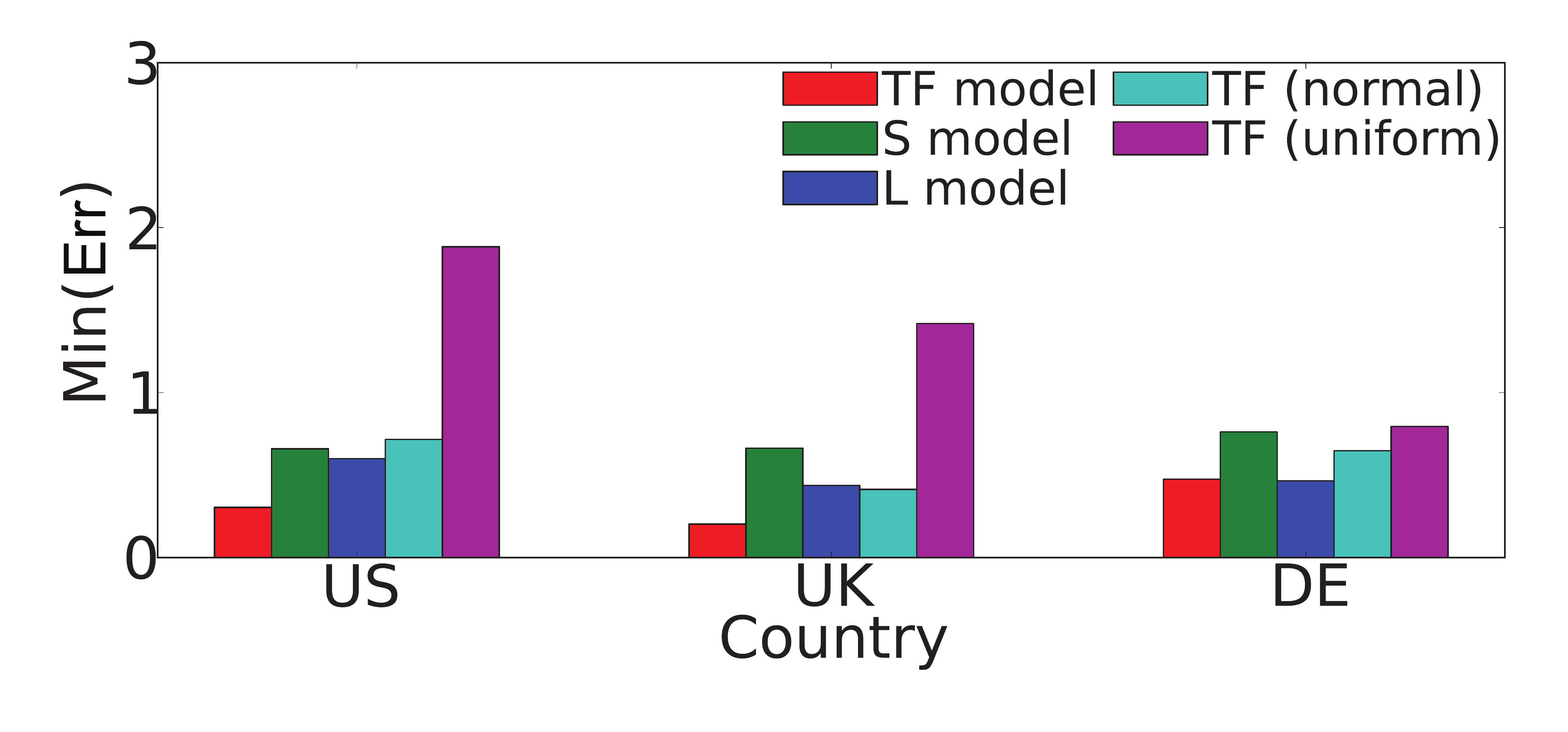}
\caption{\textbf{Comparison of different models.} The minimal values of the error $\text{Err}$ for the TF model, the two null models: spatial (S model) or linking (L model), and the TF model with normally or uniformly distributed travel distances.}
\label{fig_summary}
\end{figure}

%%%
\subsection{Sensitivity of the TF model to the parameters and its modifications}
\label{sec:sensitivity}

%%%%%%%%%%%%%%%%%%%%%%%%%%%%%%%%%%%%%%%%%%%%%%%%%%%%%%%%%%%%%%%%%%%%%%
\begin{figure*}
\centering
\includegraphics[width=17.35cm]{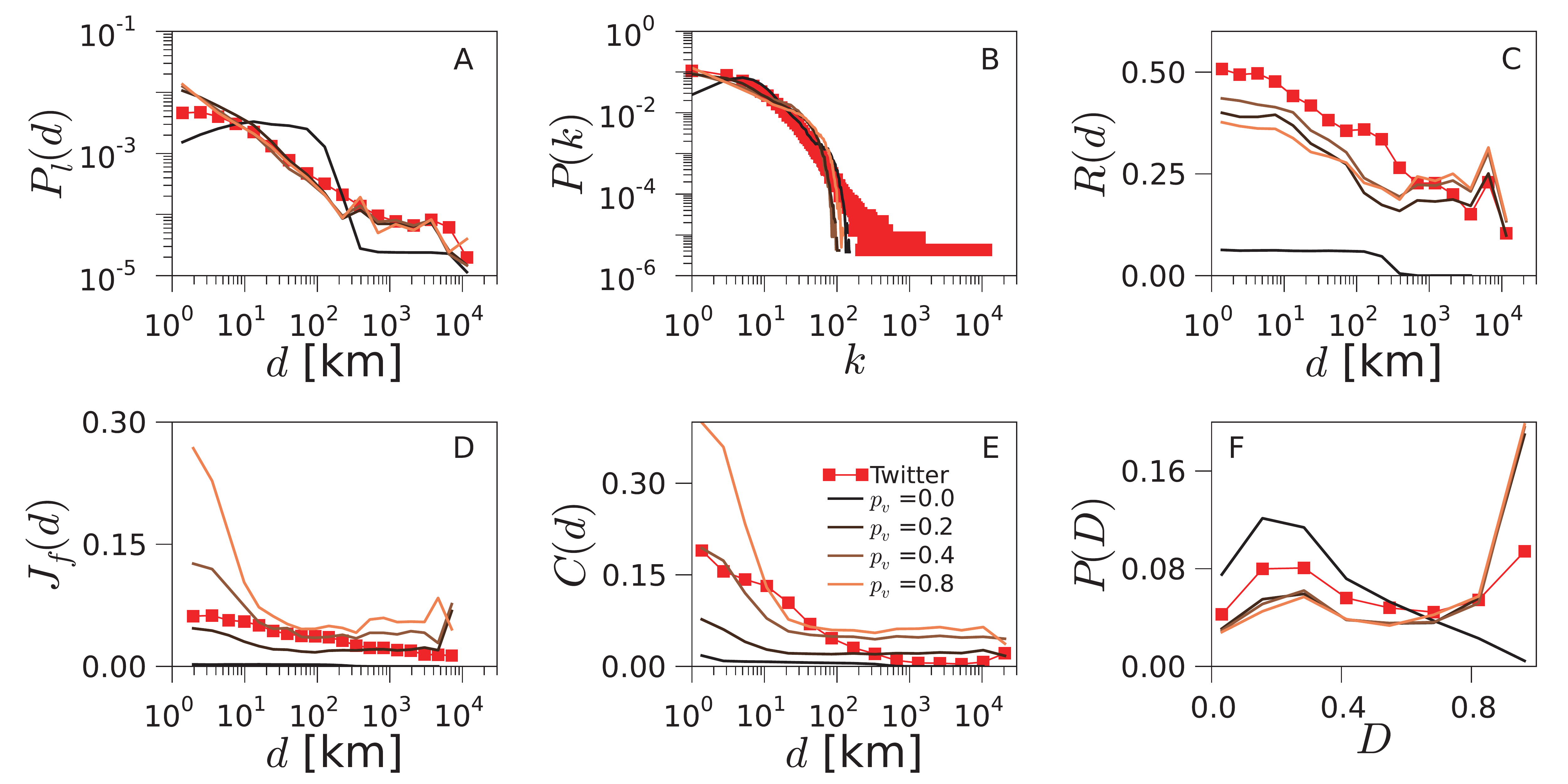}
\caption{\textbf{Impact of $p_\text{v}$ on the TF model.} We change the value of $p_\text{v}$ while keeping $p_\text{c}$ fixed to the optimal value. Note that this corresponds to an exploration of the parameter space along the vertical line crossing the minimum of $\text{Err}$ as plotted in Figure~3 for the US. Corresponding results for the UK and Germany can be found in Supplementary Figures~\ref{supfig_5} and \ref{supfig_6}.}
\label{fig_impact_fv}
\end{figure*}

%%%%%%%%%%%%%%%%%%%%%%%%%%%%%%%%%%%%%%%%%%%%%%%%%%%%%
\begin{figure*}
\centering
\includegraphics[width=17.35cm]{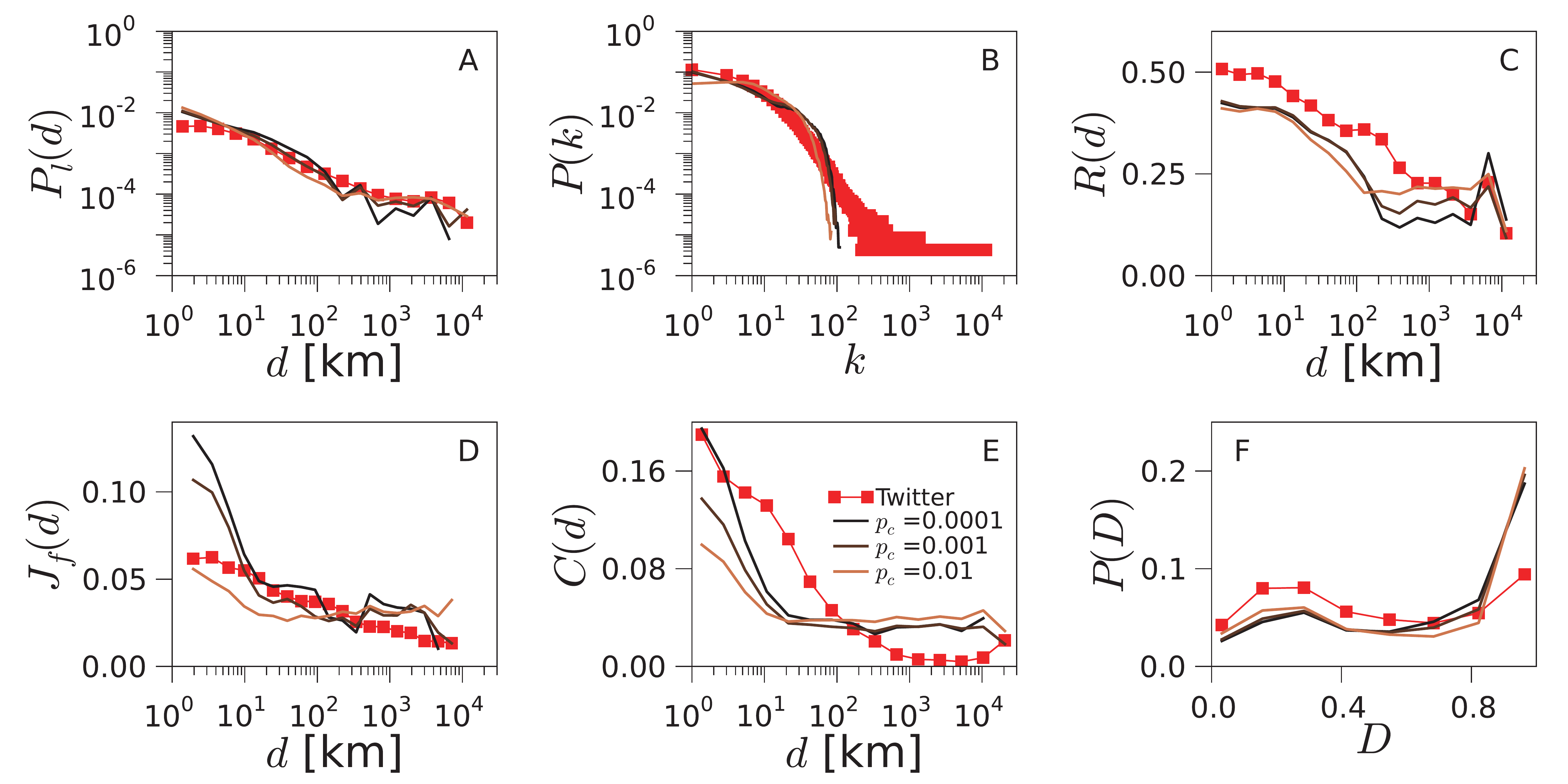}
\caption{\textbf{Impact of $p_\text{c}$ on the TF model.} We change the value of $p_\text{c}$ while keeping $p_\text{v}$ fixed to its optimal value. Note that this corresponds to an exploration of the parameter space along the horizontal line crossing the minimum of $\text{Err}$ as plotted in Figure~3 for the US. Corresponding results for the UK and Germany can be found in Supplementary Figures~\ref{supfig_7} and \ref{supfig_8}.}
\label{fig_impact_rc}
\end{figure*}

The results presented so far have been obtained at the optimal values of $p_\text{v}$ and $p_\text{c}$. The question remains, however, of how robust these results are to changes in the values of the parameters. In Figure~7, we report the effect of varying $p_\text{v}$ while $p_\text{c}$ is maintained constant in its optimal value. The linking probability $P_\text{l}\left(d\right)$ loses its power-law shape for very low values of $p_\text{v}$, marking the limit in which random mobility is the main mechanism for the agents' traveling in detriment of friend visits. In this case, most of the links are created due to encounters occurring in nearby locations or are random connections, and so the distribution of triangles disparity $P\left(D\right)$ loses its bimodal shape. Furthermore, the friend visits provide opportunities to reciprocate the connections. This is why for extremely low values of $p_\text{v}$, the reciprocity $R\left(d\right)$ is close to zero. Towards the other limit, {\it i.e.}, $p_\text{v} \to 1$ the social overlap $J_\text{f}(d)$ and the triangle-closing probability $C\left(d\right)$ steadily increase. In this limit, the linking probability $P_\text{l}\left(d\right)$, the reciprocity $R\left(d\right)$ and the distribution of triangles disparity $P\left(D\right)$ recuperate their shapes of the optimum.

Figure~8 explores the impact of varying $p_\text{c}$ while $p_\text{v}$ is fixed to its optimal value. The effect of $p_\text{c}$ on $J_\text{f}(d)$ and $C(d)$ is the opposite to that of $p_\text{v}$: these metrics decrease at all distances with increasing $p_\text{c}$. The reason for this is that visits to friends are the main forces behind the creation of new triads and the subsequent closure of triangles. Note that the more connections are created randomly (higher $p_\text{c}$), the less links will be a result of friend visits. We will expose and describe in detail the interplay between these two mechanisms in the mean-field calculations.

A possible variation of the TF model consists of eliminating friend visits or random connections ({\it i.e.}, setting $p_\text{v}$ or $p_\text{c}$ to $0$). This prevents the model from producing networks with characteristics comparable to the real ones in all the cases, leading to increase in $\text{Err}$ of around $0.5$. Interestingly, the model results are quite robust to variations in the update rules, the random connection mechanism, the connecting rules in each agent neighborhood and the variants in the way users visit friends. These variations lead to changes in $\text{Err}$ smaller than $0.1$. A detailed discussion of the results with different model variants is included in Appendix B.

%%%
\subsection{Mean-field approach}
\label{sec:mf}

In this section, we consider the L model, introduced in Section~\ref{sec:insights}, to gain some analytical insights on the mechanisms ruling the final network structure. Although this model is a simplified version of the TF model, the results of the simulations yield a relatively low value of $\text{Err}$ (Figures~6, and Supplementary Figures~\ref{supfig_9} and \ref{supfig_10} of Appendix B). We write the equations for the time evolution of the properties of the network and solve them numerically. Among all the properties, we focus on the average clustering coefficient $C$, the overall reciprocity $R$ and the degree distribution $P(k)$.

The clustering coefficient is defined as a ratio of all the closed triads to all triads existing in the network, {\it i.e.}, $C=\Delta/\Lambda$. The number of triads $\Lambda$ can be calculated knowing the degree distribution. The number of closed triads $\Delta$ in the L model grows with time mostly due to the friend visits mechanism. A triangle is formed every time two friends of the same hosting agent meet in the host's place and decide to connect. Note that an undirected triangle corresponds to $3$ undirected closed triads. Assuming that the contribution of random links is negligible, the evolution of the number of closed triads is described by
\begin{equation}
\frac{d\Delta}{dt} = 3\, N\left(k>0\right)\, \left(1-\left(1-p\right)^2\right) \left(1-C\right)\,M\,S, \label{eq_triangles}
\end{equation}
where $k=\left(k^\text{in}+k^\text{out}\right)/2$, meaning that we do not distinguish between in-degree and out-degree;
%where $k$, in this case, represents the average degree of the agents. Taking into account that the connections are directed, the agents have a certain in-degree $k^\text{in}$ and out-degree $k^\text{out}$. In the calculations, we do not distinguish between the two assuming that $k=\left(k^\text{in}+k^\text{out}\right)/2$. 
$N\left(k>0\right)$ represents the number of nodes with the degree higher than $0$, {\it i.e.}, the number of potential hosts, $M$ is an estimate of the lower bound for the number of triangles closed by one closing link $M=1+C^2\, \left(\frac{2}{1+R}\,k-2\right)$. Finally, $S$ is the expected number of encounters per host, which can be calculated as
\begin{align}
S= &\sum_{k=2}^{\infty}\, \frac{N\left(k\right)}{N} \nonumber\\
 & \times \, \sum_{i=2}^{k}
\left(\frac{p_\text{v}}{\langle k\rangle}\right)^i \,
\left(1-\frac{p_\text{v}}{\langle k\rangle}\right)^{k-i}
\binom{k}{i}\, \binom{i}{2},
\end{align}
where $N(k)$ is the number of nodes with a given degree $k$ in the network. Finally, note that the above definition of degree and the one obtained from symmetrizing directed networks (used in previous sections) are related by a proportionality factor $k = k_\text{sym}(1+R)/2$.

The reciprocity of connections $R$ can be expressed as $R=L_\text{p}/(L_\text{p}+2L_\text{s})$, where $L_\text{p}$ is the number of reciprocated links, $L_\text{s}$ is the number of non-reciprocated links and the total number of links $L=L_\text{s}+L_\text{p}$. The numbers of links evolve as
\begin{eqnarray}
\frac{dL_\text{p}}{dt} &=& 2\, N(k>0)\, \nonumber\\
 && \times \, \{ p_\text{rec} + p^2\,(1-C)\,S + p\, (1-R)\, C\,S \}, \label{eq_links_p}\\
\frac{dL_\text{s}}{dt} &=& p_\text{c}\,N + \frac{1}{3M}\,\frac{d\Delta}{dt} - \frac{1}{2}\,\frac{dL_\text{p}}{dt}, \label{eq_links_s}
\end{eqnarray}
where $p_\text{rec}=p\,p_\text{v}\,\left(1-p_\text{v}\right)\,\left(1-R\right)$ corresponds to the probability that an agent visiting a neighbor gets her connection reciprocated (their connection is initially single directional).
As can be seen, $\Delta$, $L_\text{p}$ and $L_\text{s}$ are mutually dependent.

%%%%%%%%%%%%%%%%%%%%%%%%%%%%%%%%%%%%%%%%
\begin{figure*}
\centering
\includegraphics[width=17.35cm]{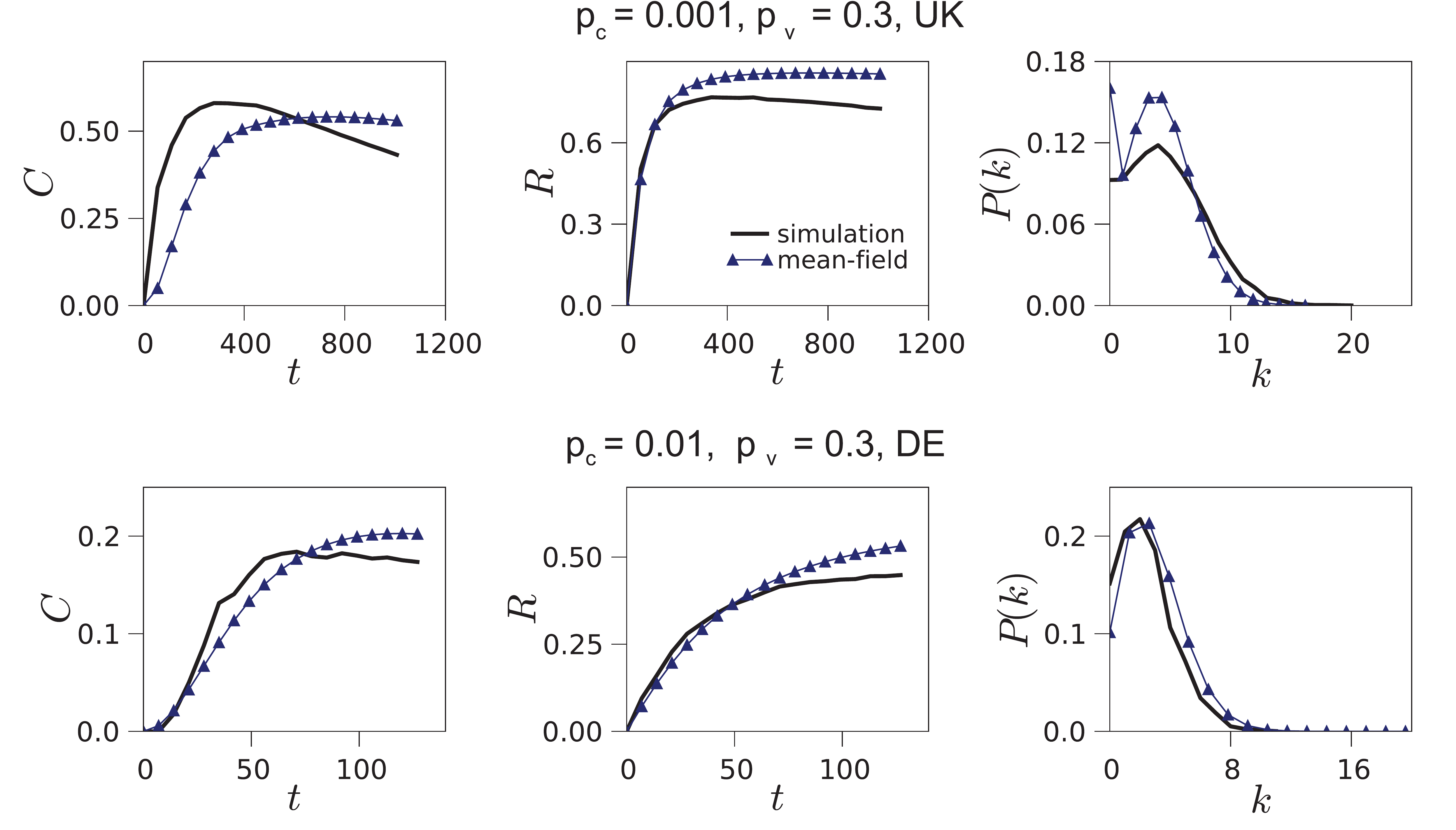}
\caption{\textbf{Mean-field approximation.} Predictions of the analysis versus results of the simulation of the L model for the clustering coefficient $C$, the reciprocity $R$ and the degree distribution $P(k)$. In this case, we are taking the users from the UK and Germany because their lower numbers facilitate the numerical integration of the Equations~\ref{eq_triangles}, \ref{eq_links_p}, \ref{eq_links_s}, \ref{eq_degdistr3} and~\ref{eq_degdistr4}.}
\label{fig_mfapprox}
\end{figure*}

To calculate the degree distribution $P\left(k\right)$, we estimate the probability $p_\text{con}$ of a node to increase its degree by one unit in the current time step due to multiple encounters with friends of her friends
\begin{equation}
p_\text{con} = \sum_{k'=2}^{\infty} \, \frac{k'\, N\left(k'\right)}{\langle k\rangle N} \,
\binom{k'-1}{2}\, p_\text{c}^2\, \left(1-p_\text{c}\right)^{k'-2},
\label{pk}
\end{equation}
where $p_\text{c} = p \,p_\text{v}/\langle k\rangle\, \left(1-\left(1+R\right)/2C\right)$. In the L model, however, every node can increase its degree by multiple links in each time step. For simplicity, we neglect higher order terms induced by the possibility of creating multiple links. Moreover, we note that Equation~(\ref{pk}) is a good estimate if there is not a strong correlation between node degrees. The number of nodes of certain degree $k$ is given by
\begin{eqnarray}
k>1: \hspace{5mm}
\frac{dN\left(k\right)}{dt} &=& p_\text{inc}\, \left(N\left(k-1\right)-N\left(k\right)\right), \label{eq_degdistr1} \nonumber\\
\frac{dN\left(1\right)}{dt} &=& p_\text{c} \,N\left(0\right) -p_\text{inc}\,N\left(1\right) + p_\text{rec}\, N_\text{s}\left(0\right), \label{eq_degdistr2}\nonumber \\
\frac{dN\left(0\right)}{dt} &=& - p_\text{c}\, N\left(0\right) - p_\text{rec}\, N_\text{s}\left(0\right), \label{eq_degdistr3}
\end{eqnarray}
where $p_\text{inc} = p_\text{c} + p_\text{rec}/2 + p_\text{v}\,p_\text{con}$ is an estimate of the probability that the node degree increases, $N_\text{s}(0)$ is the number of nodes with 0 out-degree and non-zero in-degree. Such nodes are important because their connection can be easily reciprocated as a result of a friend visit. However, these nodes are not counted directly into $N\left(1\right)$, and so a correction is needed to account for them explicitly, as in Equation~(\ref{eq_degdistr3}). The number of such nodes can be calculated as
\begin{equation}
\frac{dN_\text{s}(0)}{dt} = p_\text{c} N\left(0\right) - p_\text{rec} N_\text{s}\left(0\right). \label{eq_degdistr4}
\end{equation}

The numerical solution of this set of equations describing the evolution of the L model is shown in Figure~9. The equations accurately predict the dynamics of the clustering coefficient $C$, the reciprocity $R$ and the degree distribution $P(k)$ for certain values of the parameters ({\it i.e.}, for medium and high values of $p_\text{c}$, as in the lower plots of Figure~9). The approximation yields slightly worse results when the number of random connections is small in comparison with the number of connections created due to friend visits ({\it i.e.}, for low values of $p_\text{c}$, as in the upper plots of Figure~9B). In the latter case, neither the degree distribution is well approximated, probably due to the degree-degree correlations introduced through the friend visit mechanism.

The mean-field analysis of the L model shows that the friend visiting mechanism is a direct cause of triangle closure and link reciprocity. Equation~\ref{eq_triangles}, which estimates the growth of the number of triangles in the network, accounts only for the friend visiting mechanism; yet it approximates closely the value of the clustering coefficient, also when $p_\text{c}$, which controls the mechanism of random connections, is high. Similarly, Equation~\ref{eq_links_p}, which estimates the growth of the number of reciprocated connections, accounts for the friend visiting mechanism and approximates well the value of reciprocity.

\section{Discussion}

% ===== intro, two parameters, reproduces data, insights
We introduce a model that couples human mobility and link creation in social networks. The aim is to characterize the relation between network topology and geography observed in empirical online networks. The model has two free parameters $p_\text{c}$ and $p_\text{v}$ but, despite its simplicity, it is able to reproduce a good number of geo-social features observed in real data at a country level. Comparing the TF model with simplified null models, we find that the coupling of geography and structure through a realistic mobility mechanism produces significantly more realistic social networks than the uncoupled models.

% ===== relational and proximity mechanisms
% ===== main mechanism at play
Social links in our model are formed mostly with relational (due to triadic closure), and proximity (through spatio-temporal coincidences) mechanisms~\cite{rivera10}. Visiting friends helps to reinforce the existing relations and favors the closure of triads with particular properties regarding the distance balance of their edges. Random link creation accounts for online acquaintances or for historical face-to-face encounters as individuals move their residence from one city to another. Finally, individual random mobility allows the agents to explore new locations.\footnote{We expect that in our model the number of unique locations visited over time grow linearly in time, due to the fact that with constant probability $(1-p_v)$ an agent jumps to a new location, as opposed to slower growth reported in \cite{Song2010Modelling}. We leave the exploration of temporal aspects of our model for the future research.} Our results show that by establishing an appropriate balance between friend visits and random link creation, the model can reproduce the main features of online social networks, e.g., we show that $10\%-30\%$ of the mobility has to be directed towards existing friends. We demonstrate that these are the fundamental mechanisms at play in the model.

% ===== use cases
The TF model is generic and functional for different datasets. Human mobility driven by social ties has impact on the modeling of disease spreading, and may improve its predictions. This model can also be used in simulations of processes that involve social networks and geography, e.g., simulations of opinion formation, language evolution, or responses of a population to extreme events. Moreover, it can also be helpful to design network benchmarks with realistic geo-social properties to test, for instance, the scalability of technical solutions in social online networks related to geography of its physical infrastructure.

\subsection{Ethics Statement}

The data analyzed are publicly available as they come from public online social sites or data repositories (Twitter, Gowalla and Brightkite). Since our analysis relies on statistical features and not on single cases, any private information about users had been removed and the analysis was performed on anonymized datasets.

%% Do NOT remove this, even if you are not including acknowledgments
\section{Acknowledgments}

We would like to warmly thank Luis F. Lafuerza for helpful discussions on the analytical treatment of the model. 
% This paragraph below has to be copied & pasted in the funding part of the submission and then commented here.
%We have received partial financial support from the Spanish Ministry of Economy (MINECO) and FEDER (EU) under project MODASS (FIS2011-24785), and from the EU Commission through projects EUNOIA, INSIGHT and LASAGNE. PAG acknowledges funding from the JAE-Predoc program of CSIC and JJR from the Ram\'on y Cajal program of MINECO. BG was partially supported by the French ANR project HarMS-flu (ANR-12-MONU-0018).

\appendix

\section{Supplementary Figures}

In this Appendix, we present Supplementary Figures that reproduce the figures of the main text for the United Kingdom and Germany.

\begin{supfigure*}[htb]
\centering
 \includegraphics[width=0.90\textwidth]{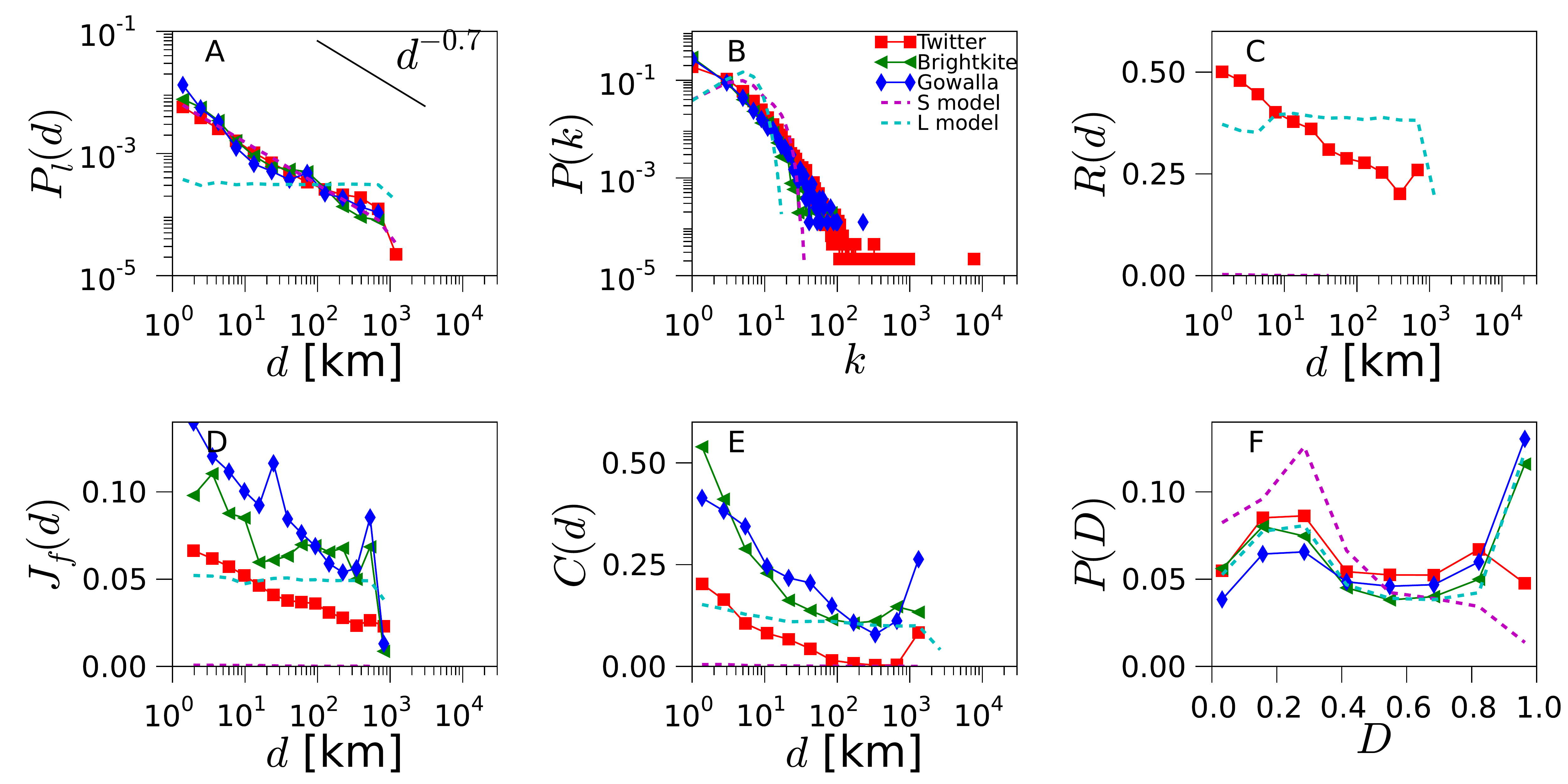}
\caption{
\textbf{Network geo-social properties.} Various statistical network properties are plotted for the data obtained from Twitter (red squares), Gowalla (blue diamonds), Brightkite (green triangles) and the null models (dashed lines), for the UK.
}
\label{supfig_1}
\end{supfigure*}

\begin{supfigure*}[htb]
\centering
 \includegraphics[width=0.90\textwidth]{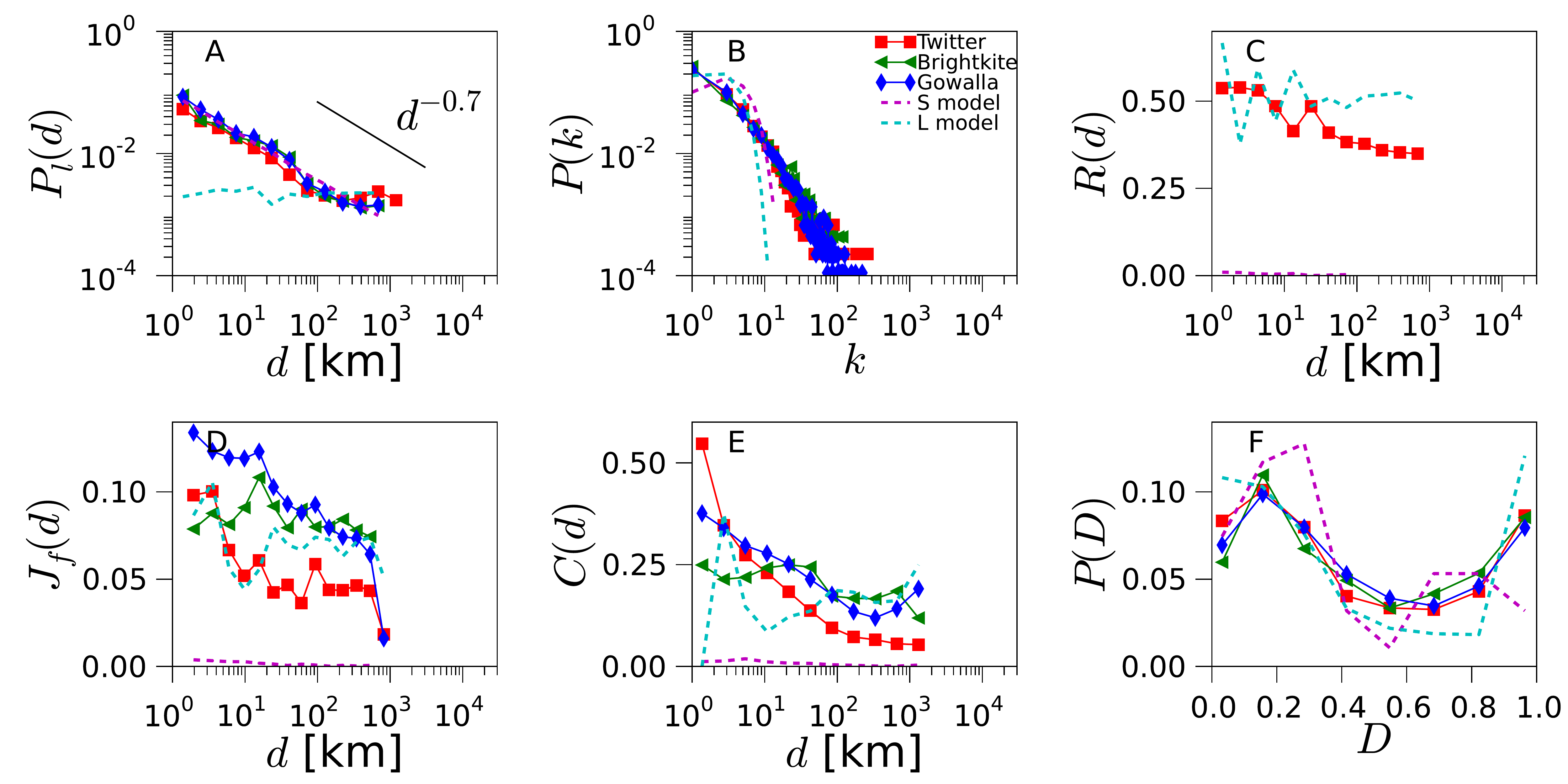}
\caption{
\textbf{Network geo-social properties.} Various statistical network properties are plotted for the data obtained from Twitter (red squares), Gowalla (blue diamonds), Brightkite (green triangles) and the null models (dashed lines), for Germany.
}
\label{supfig_2}
\end{supfigure*}

\begin{supfigure*}[p]
\centering
 \includegraphics[width=0.90\textwidth]{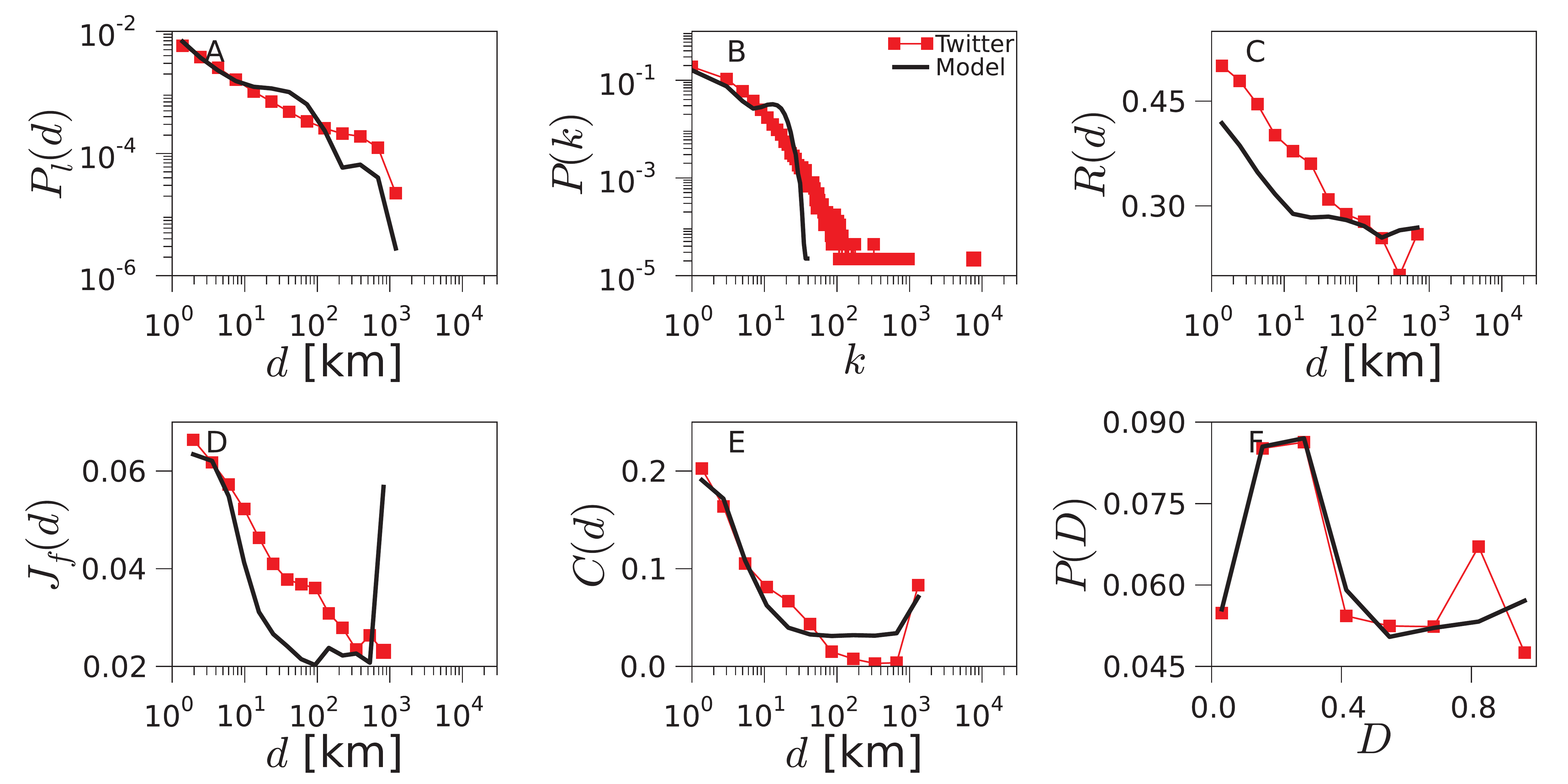}
\caption{
\textbf{Geo-social properties of the model networks.} Various statistical properties are plotted for the networks obtained from Twitter data (red squares) and from simulation of the TF model (black line) for the UK.
}
\label{supfig_3}
\end{supfigure*}

\begin{supfigure*}[p]
\centering
 \includegraphics[width=0.90\textwidth]{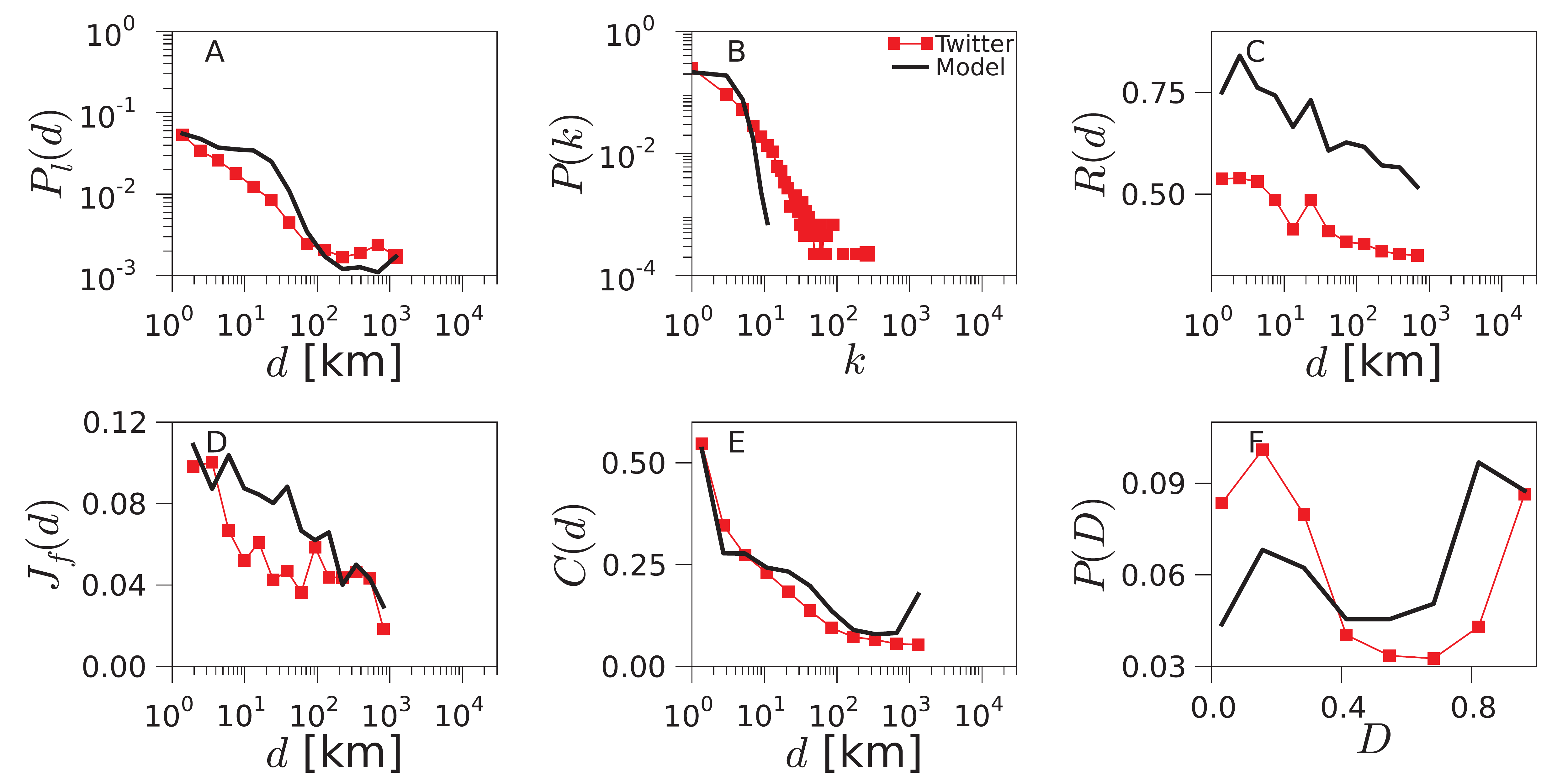}
\caption{
\textbf{Geo-social properties of the model networks.} Various statistical properties are plotted for the networks obtained from Twitter data (red squares) and from simulation of the TF model (black line) for Germany.
}
\label{supfig_4}
\end{supfigure*}

\begin{supfigure*}[p]
\centering
 \includegraphics[width=0.90\textwidth]{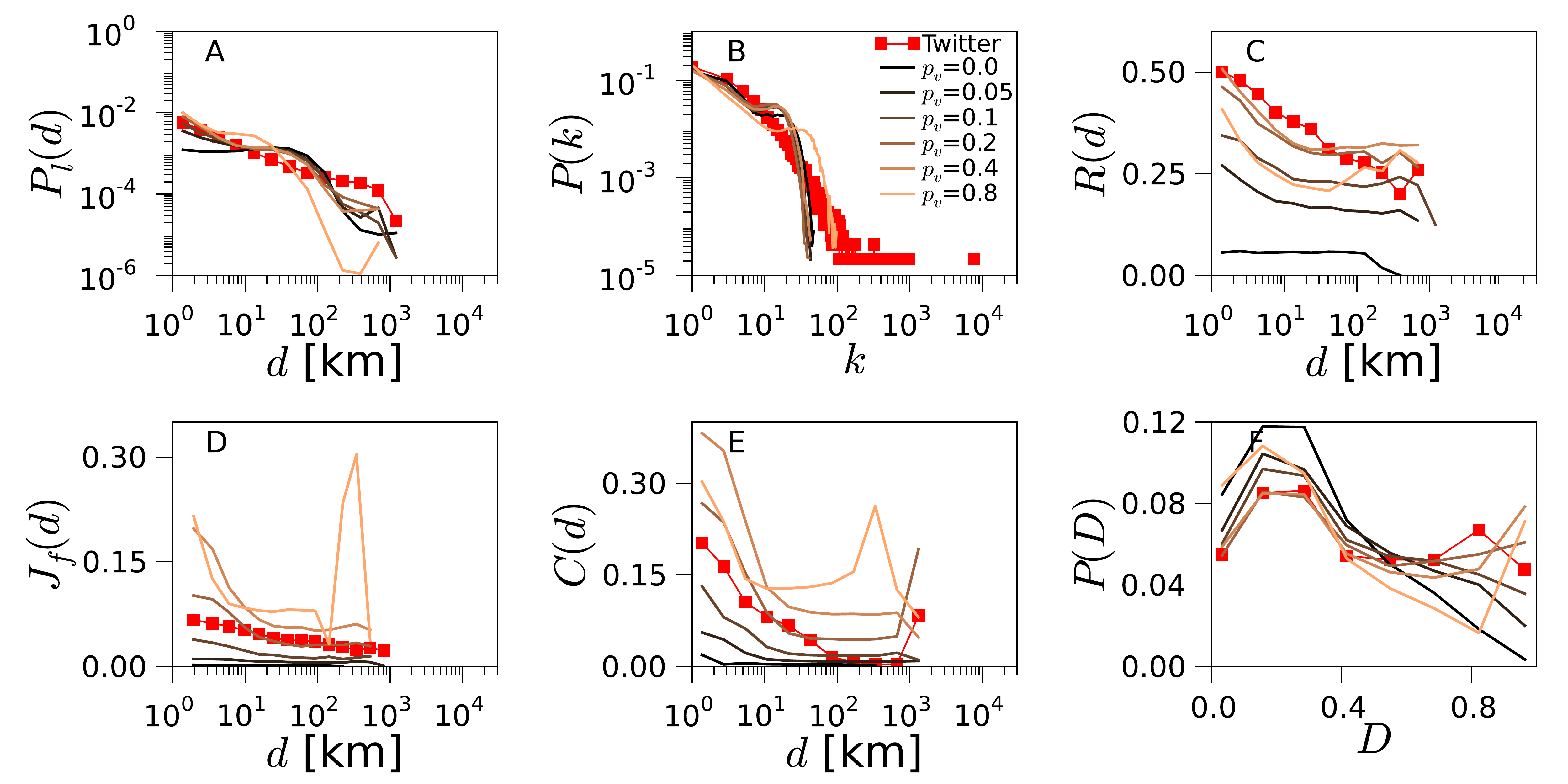}
\caption{
\textbf{Impact of $p_\text{v}$ on the TF model.} We change the value of $p_\text{v}$ while keeping $p_\text{c}$ fixed to the optimal value. Note that this corresponds to an exploration of the parameter space along the vertical line crossing the minimum of $\text{Err}$ as plotted in Figure~3 for the UK.
}
\label{supfig_5}
\end{supfigure*}

\begin{supfigure*}[p]
\centering
 \includegraphics[width=0.90\textwidth]{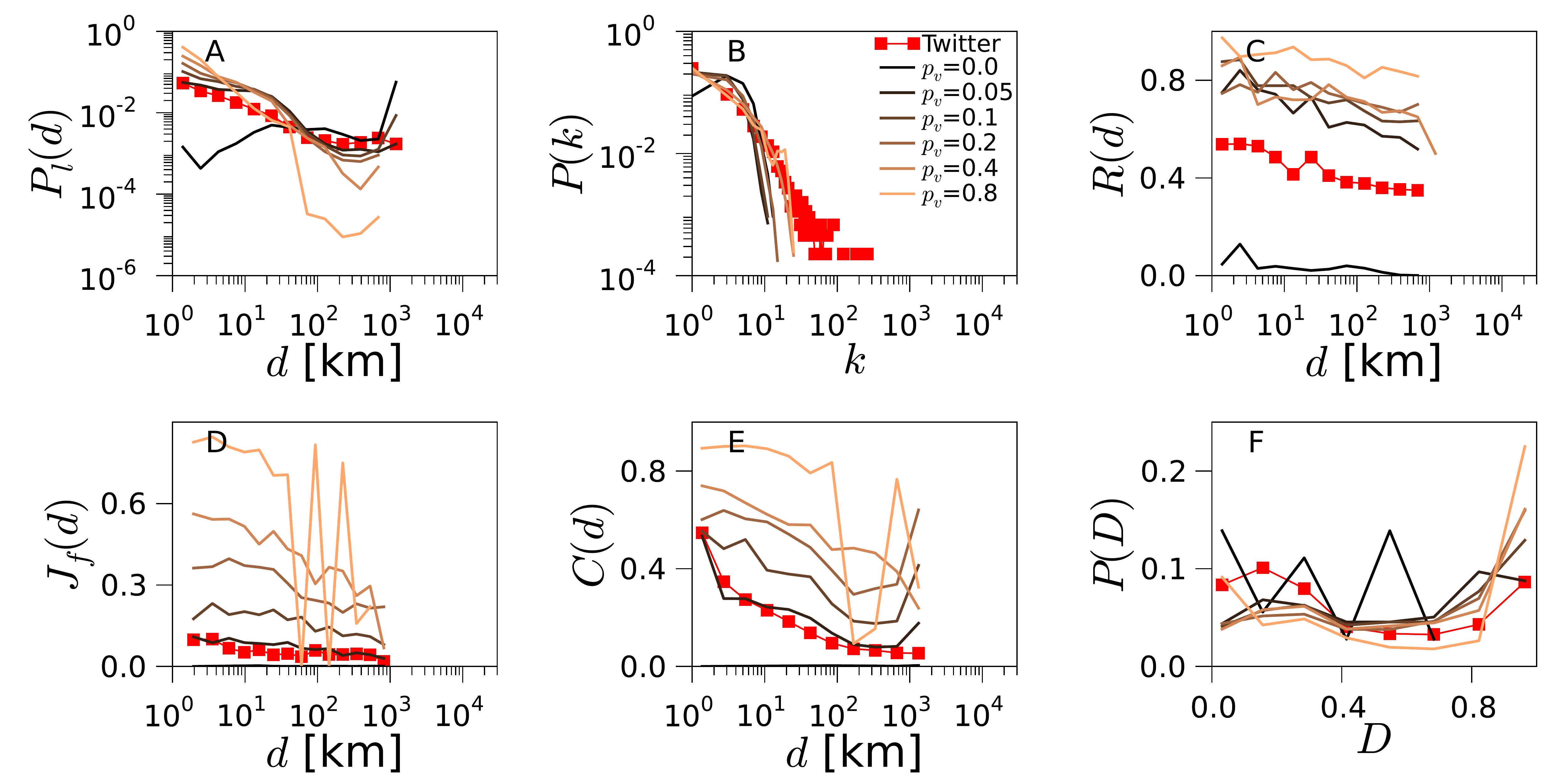}
\caption{
\textbf{Impact of $p_\text{v}$ on the TF model.} We change the value of $p_\text{v}$ while keeping $p_\text{c}$ fixed to the optimal value. Note that this corresponds to an exploration of the parameter space along the vertical line crossing the minimum of $\text{Err}$ as plotted in Figure~3 for Germany.
}
\label{supfig_6}
\end{supfigure*}

\begin{supfigure*}[p]
\centering
 \includegraphics[width=0.90\textwidth]{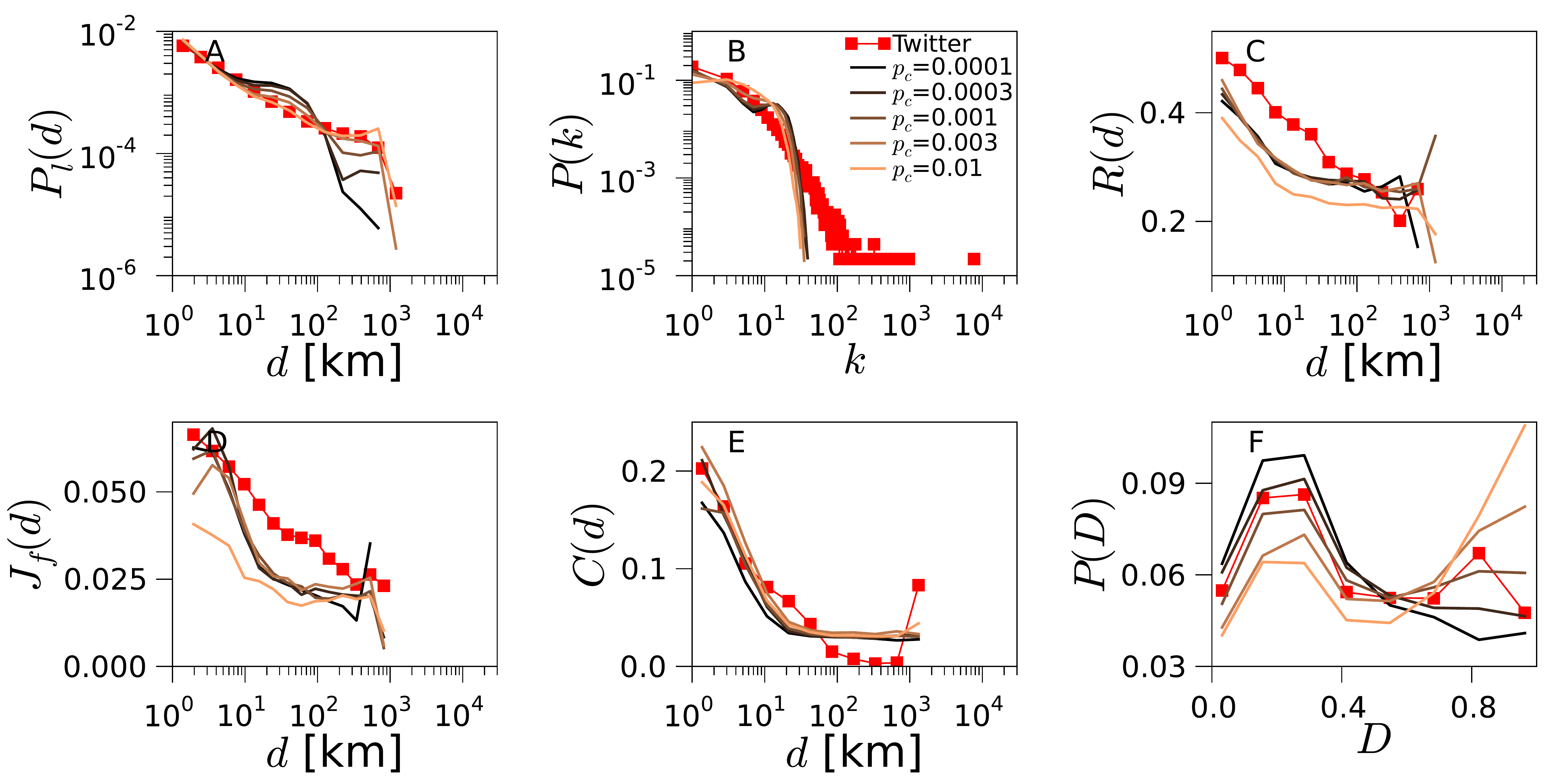}
\caption{
\textbf{Impact of $p_\text{c}$ on the TF model.} We change the value of $p_\text{c}$ while keeping $p_\text{v}$ fixed to its optimal value. Note that this corresponds to an exploration of the parameter space along the horizontal line crossing the minimum of $\text{Err}$ as plotted in Figure~3 for the UK.
}
\label{supfig_7}
\end{supfigure*}

\begin{supfigure*}[p]
\centering
 \includegraphics[width=0.90\textwidth]{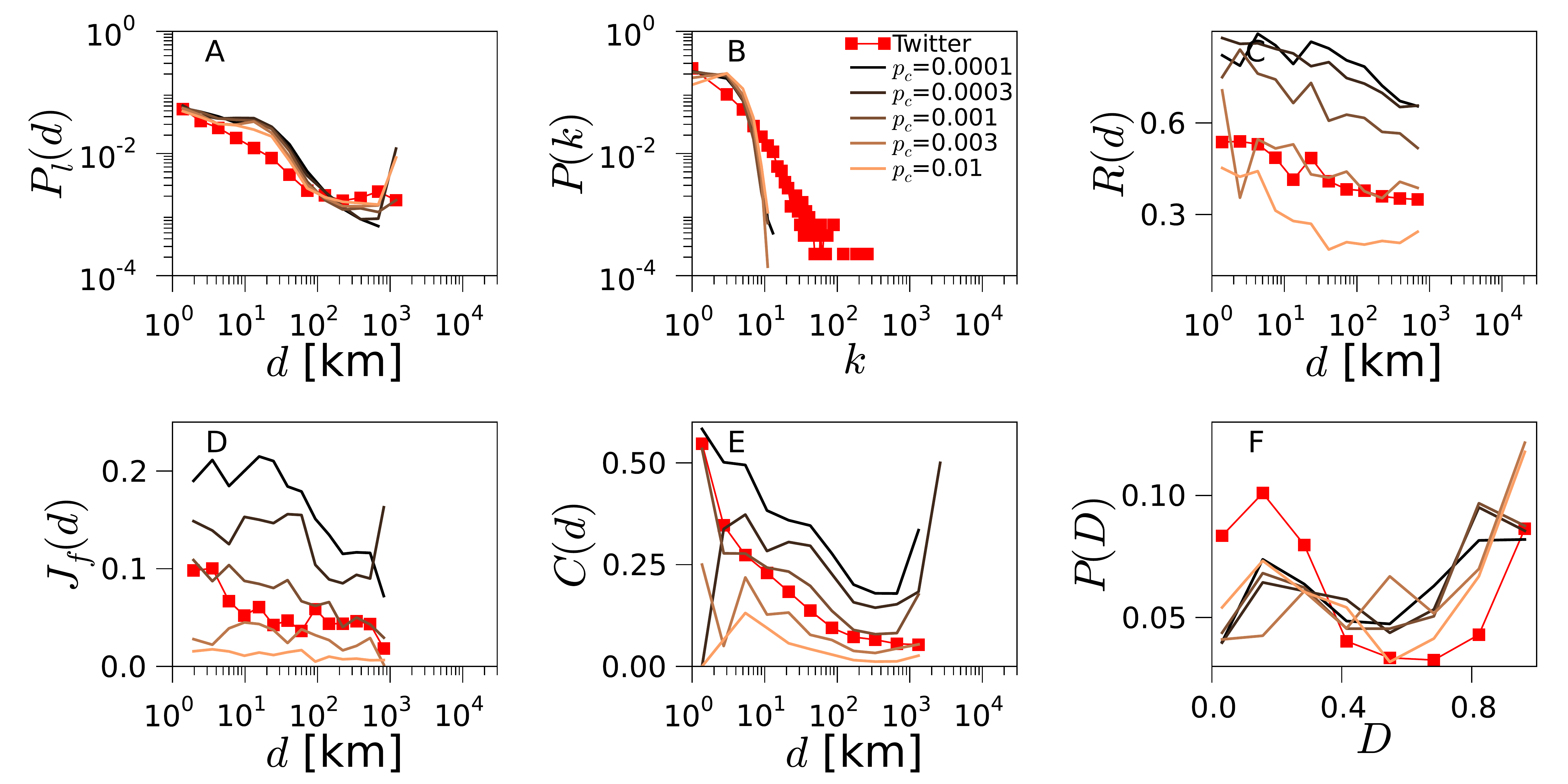}
\caption{
\textbf{Impact of $p_\text{c}$ on the TF model.} We change the value of $p_\text{c}$ while keeping $p_\text{v}$ fixed to its optimal value. Note that this corresponds to an exploration of the parameter space along the horizontal line crossing the minimum of $\text{Err}$ as plotted in Figure~3 for Germany.
}
\label{supfig_8}
\end{supfigure*}

\FloatBarrier

\section{Variants of the TF model}

In this Appendix, we consider several variants of the TF model and the L model and evaluate their results. We describe a total of $36$ variants marked with different colors in the tables in Supplementary Figures~\ref{supfig_9} and~\ref{supfig_10}. For each variant we explore the space of the parameters $p_\text{v}$ and $p_\text{c}$. We run the models for $p_\text{v}$ from the set $\{0, 0.1, 0.2, 0.3, 0.4, 0.5, 0.6, 0.7\}$ and $p_\text{c}$ from $\{0, 0.0003, 0.001, 0.003, 0.01, 0.03\}$, yielding in total $48$ parameter combinations. For each of the model variants, we find the parameters that minimize the fitting error $\text{Err}$. We plot its value in Supplementary Figures~\ref{supfig_9} and~\ref{supfig_10}. In the following paragraphs, we describe in detail each of the variants and its results.

\begin{supfigure*}[tbph]
\centering
 \includegraphics[width=0.90\textwidth]{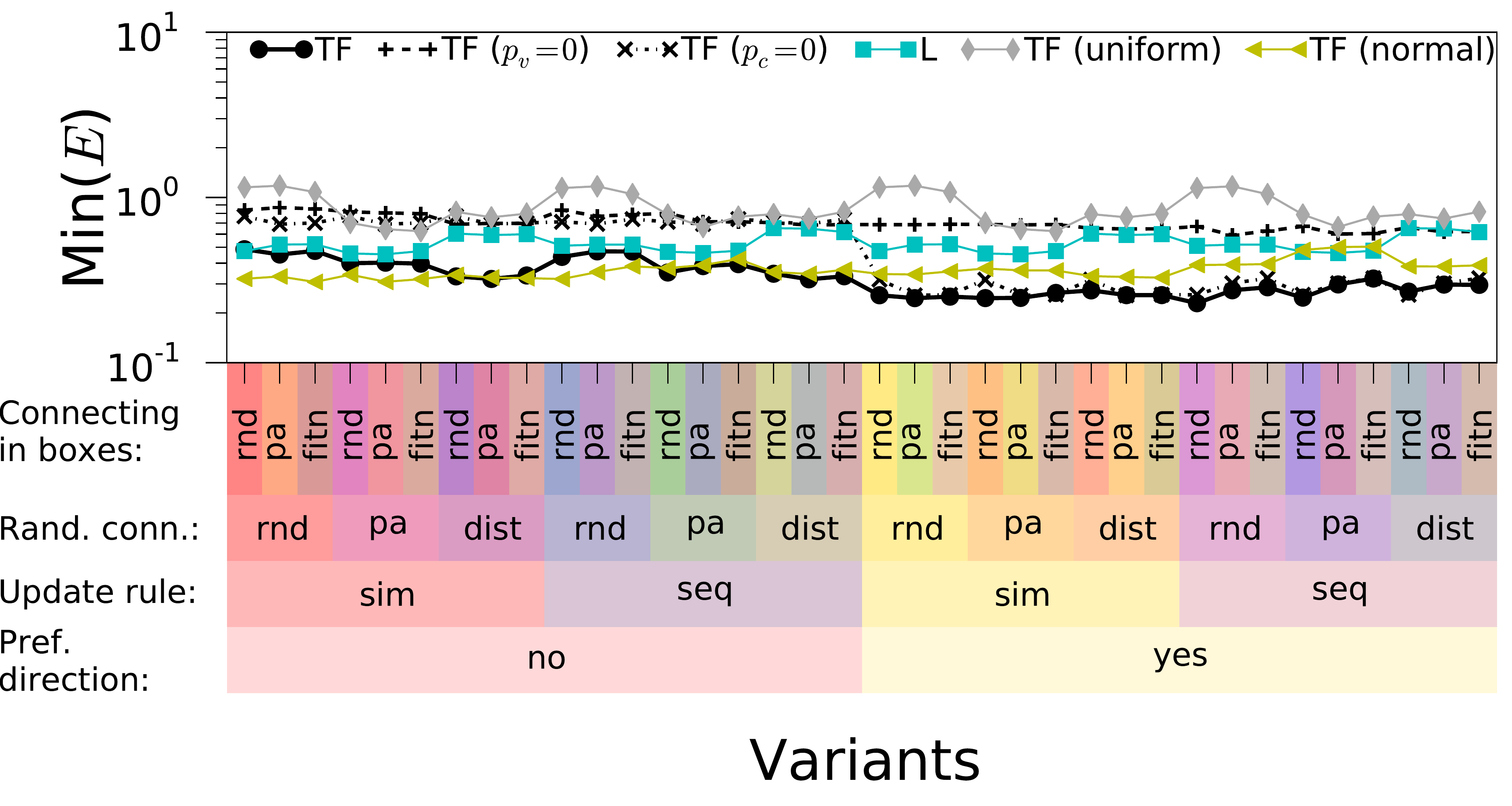}
\caption{
\textbf{The model variants.} Values of the fitting error $\text{Err}$ for the UK for the variants of the following models: the TF model, the TF model with $p_\text{v}=0$, the TF model with $p_\text{c}=0$, the L model and the TF model with uniformly or normally distributed jumps. The default variant described in the manuscript is marked with the the red rectangle.
}
\label{supfig_9}
\end{supfigure*}

\begin{supfigure*}[tbph]
\centering
\includegraphics[width=0.90\textwidth]{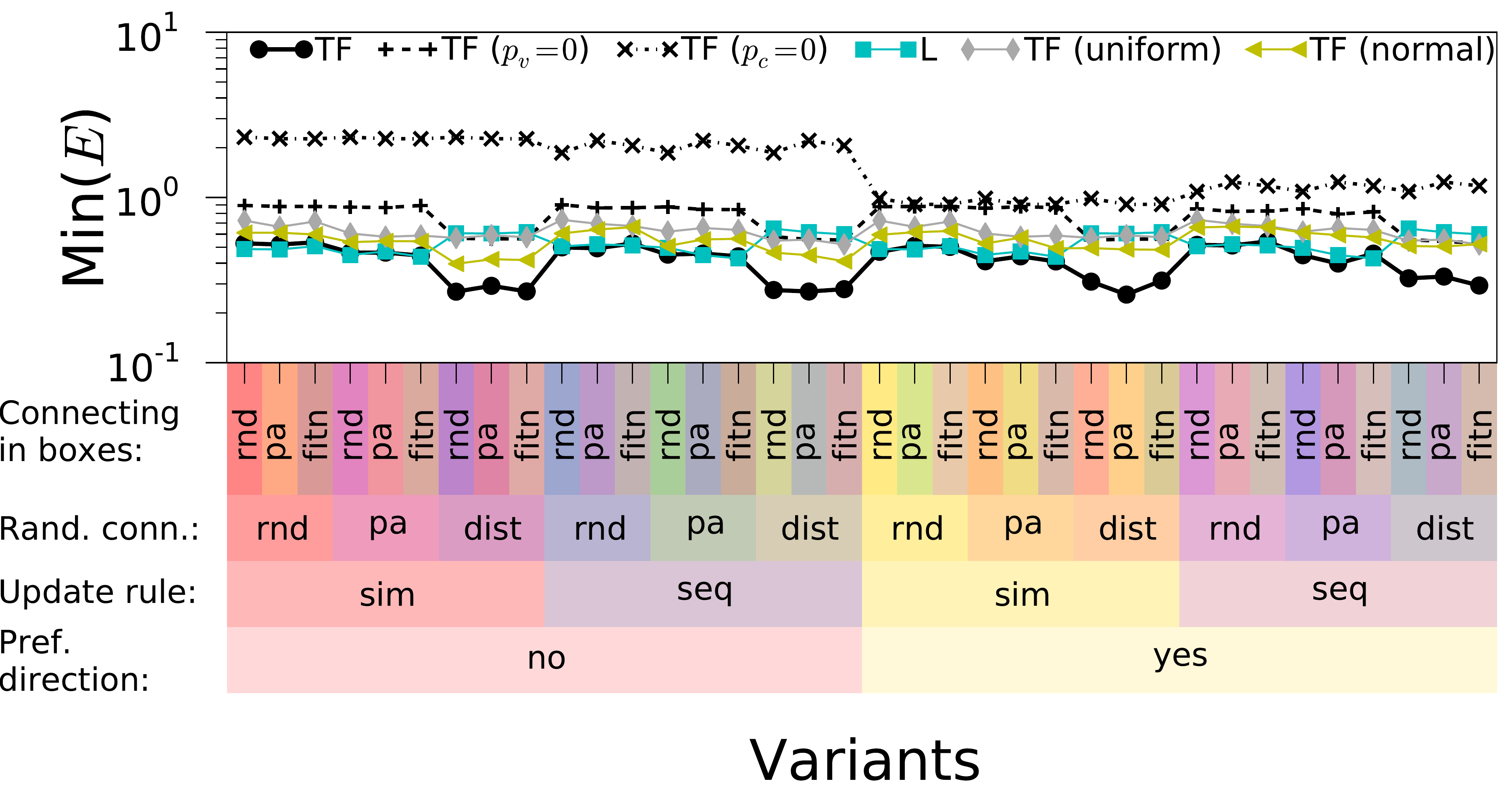}
\caption{
\textbf{The model variants.} Values of the fitting error $\text{Err}$ for Germany for the variants of the models as in Supplementary Figure~\ref{supfig_9}.
}
\label{supfig_10}
\end{supfigure*}

% mobility models
First, we modify the jump size distribution to understand its impact on the geo-social properties. We consider the following cases: the default power-law jumps with exponent $1.55$, a minimal jump length of $1\,$km and a cutoff at $100\,$km, as in~\cite{Song2010Modelling} (the TF model), uniformly distributed random jumps up to $100\,$km (TF-uniform), normally distributed jumps with standard deviation of $1\,$km (TF-normal), and no jumps to new locations at all (the L model). 
We plot the minimal fitting error of these cases in Supplementary Figures~\ref{supfig_9} and~\ref{supfig_10} using different colors of the curves.
The default power-law jumps show the best results with the lowest error for most of the variants. The normal distribution and the L model tend to perform considerably worse. The highly unrealistic uniform jumps understandingly provide the worst results and the highest error values for almost all variants.

To assess the role of friend visits and random connections we turn on and off these two components by setting to zero the corresponding parameters $p_\text{v}$ and $p_\text{c}$. We plot the results in Supplementary Figures~\ref{supfig_9} and~\ref{supfig_10} with dashed and dotted lines. We observe significantly higher error values whenever one of these two components is turned off, for most of the model's variants, what demonstrates their importance for the TF model.

To prevent users from spreading into inhabited regions, we include in the TF model an angular preference for the jumps. Namely, the direction of each jump is chosen randomly with a probability proportional to the number of inhabitants present at the destination. Note that this does not affect in any way the length of the jumps, which is drawn independently beforehand. To estimate the population of the target area, we use the gridded population of the world \cite{SEDAC2014}. To test how this angular preference impacts the results, we consider a variation of the models without it and compare the results.
The two variants are included in the lowest row of the table in Supplementary Figures~\ref{supfig_9} and~\ref{supfig_10}. They show almost no difference in the error values for the Germany, although a systematic difference exists for the UK; the error of the variant with direction preference is consistently lower in the case of that country. The presence of the sea around the UK introduces a distorting factor for the TF model. Without the directional preference the agents freely spread over the sea independently on the geographical shape of the country, leading to unrealistic results.

% update rules
Agents' traveling and link creation can be realized in the simulation in various update orders. By default, at each time step, each agent first moves, next connects to other random agents, and then connects locally; the following agent performs the same actions in the same order, etc. We call this method sequential (``seq''). In an alternative update rule, which we call simultaneous (``sim''), first all the agents move, then all of them create random connections, and finally all of them create connections locally.
The two update rules are included in the second row, counting from the bottom, of the table in Supplementary Figures~\ref{supfig_9} and~\ref{supfig_10}. The update rules have little impact on the final networks resulting from the simulation.

% random connections
% connecting inside of boxes
In the TF and L models, the agents create with probability $p_\text{c}$ random connections. These links can be created in different ways; we consider three variants. First, each agent chooses another agent uniformly at random, what constitutes the default mechanism (``rnd''). Second, each agent randomly picks another node with probability proportional to the current degree of the node, which corresponds to the preferential attachment mechanism (``pa''). Third, the agent draws another node with probability decaying as a power-law of the distance between the two agents (``dist''), with its exponent equal to $1.4$ and the minimal distance of $0.1\,$km. The type of random connecting mechanism used is listed in the third row, counting from the bottom, of the table in Supplementary Figures~\ref{supfig_9} and~\ref{supfig_10}. In some cases, e.g., for Germany, the distant-dependent probability of creating a random link provides better results than the other variants.

We consider similar variants for the connections formed inside spatial boxes, which are created with the probability $p$. The agents can connect uniformly at random (``rnd''), with a preference for high-degree nodes (``pa''), or a preference toward the nodes with high intrinsic fitness (``fitn''). The fitness of the nodes is drawn from a power-law distribution with an exponent of $1.5$, which roughly corresponds to the distribution of the growth rates reported in \cite{Grabowicz2012Heterogeneity}. These variants are implemented in the following way.
First, we note that the number of connections created by the agent is a result of a binomial process with probability $p$ and the number of trials is equal to the number of agents that currently stay in the given spatial box. The expected number of links created in such binomial process is known, therefore, an equivalent number of connections can be created with one of the two mentioned preferential processes.
The type of connecting mechanism applied in the spatial boxes is listed in the top row of the table in Supplementary Figures~\ref{supfig_9} and~\ref{supfig_10}. There is no consistent difference in the error values between these variants. Thus, the connecting mechanism applied in the spatial boxes has little impact on the results.

We conclude that the main components of the TF models are crucial to reproduce the structure and geography of the social networks. These components include the mobility model, friend visits and random connections. The power-law mobility model tends to produce the best results. The angular preference of travels is important for countries whose geography is strongly restrained, e.g., by sea. Other modifications to the model have low or no consistent impact on the results, with the exception of the distance dependent random connections, which in certain cases consistently influence the results.

\end{document}